\documentclass[journal, 10pt]{IEEEtran}
\usepackage[T1]{fontenc}
\usepackage[latin9]{inputenc}
\usepackage{pifont}
\usepackage{float}
\usepackage{url}
\usepackage{amsmath}
\usepackage{amsthm}
\usepackage{amssymb}
\usepackage{graphicx}
\usepackage[unicode=true,
 bookmarks=true,bookmarksnumbered=true,bookmarksopen=true,bookmarksopenlevel=1,
 breaklinks=false,pdfborder={0 0 0},pdfborderstyle={},backref=false,colorlinks=false]
 {hyperref}
\hypersetup{pdftitle={Your Title},
 pdfauthor={Your Name},
 pdfpagelayout=OneColumn, pdfnewwindow=true, pdfstartview=XYZ, plainpages=false}

\makeatletter

\floatstyle{ruled}
\newfloat{algorithm}{tbp}{loa}
\providecommand{\algorithmname}{Algorithm}
\floatname{algorithm}{\protect\algorithmname}

\theoremstyle{definition}
\newtheorem{defn}{\protect\definitionname}
\theoremstyle{plain}
\newtheorem{thm}{\protect\theoremname}
\theoremstyle{plain}
\newtheorem{prop}{\protect\propositionname}

\usepackage[caption=false,font=footnotesize]{subfig}
\usepackage{algpseudocode}

\algdef{SE}{Begin}{End}{\textbf{begin}}{\textbf{end}}
\algnewcommand\algorithmicforeach{\textbf{for each}}
\algdef{S}[FOR]{ForEach}[1]{\algorithmicforeach\ #1\ \algorithmicdo}

\makeatother

\providecommand{\definitionname}{Definition}
\providecommand{\propositionname}{Proposition}
\providecommand{\theoremname}{Theorem}

\begin{document}
\title{Mechanism Design for Wireless Powered Spatial Crowdsourcing Networks}
\author{Yutao Jiao, Ping Wang, Dusit
Niyato, Bin Lin, and Dong In Kim\thanks{Copyright (c) 2019 IEEE. Personal use of this material is permitted. However, permission to use this material for any other purposes must be obtained from the IEEE by sending a request to pubs-permissions@ieee.org. 

Yutao Jiao and Dusit Niyato are with the School of Computer Science
and Engineering, Nanyang Technological University, Singapore. Ping
Wang is with the Lassonde School of Engineering, York University,
Canada. Bin Lin is with the College of Information Science and Technology,
Dalian Maritime University, China. Dong In Kim is with the School
of Information and Communication Engineering, Sungkyunkwan University,
Korea.}}
\maketitle
\begin{abstract}
	Wireless power transfer (WPT) is a promising technology to prolong the lifetime of the sensors and communication devices, i.e., workers, in completing crowdsourcing tasks by providing continuous and cost-effective energy supplies. In this paper, we propose a wireless powered spatial crowdsourcing framework which consists of two mutually dependent phases: task allocation phase and data crowdsourcing phase. In the task allocation phase, we propose a Stackelberg game based mechanism for the spatial crowdsourcing platform to efficiently allocate spatial tasks and wireless charging power to each worker. In the data crowdsourcing phase, the workers may have an incentive to misreport its real working location to improve its utility, which causes adverse effects to the spatial crowdsourcing platform. To address this issue, we present three strategyproof deployment mechanisms for the spatial crowdsourcing platform to place a mobile base station, e.g., vehicle or robot, which is responsible for transferring the wireless power and collecting the crowdsourced data. As the benchmark, we first apply the classical median mechanism and evaluate its worst-case performance. Then, we design a conventional strategyproof deployment mechanism to improve the expected utility of the spatial crowdsourcing platform under the condition that the workers' locations follow a known geographical distribution. For a more general case with only the historical location data available, we propose a deep learning based strategyproof deployment mechanism to maximize the spatial crowdsourcing platform's utility. Extensive experimental results based on synthetic and real-world datasets reveal the effectiveness of the proposed framework in allocating tasks and charging power to workers while avoiding the dishonest worker's manipulation.
\end{abstract}


\section{Introduction}

Crowdsourcing is becoming a popular paradigm which efficiently completes tasks and solves problems by aggregating information and intelligence from crowds. Integrated with advanced sensing and communication techniques, mobile devices can help complete diverse location-aware tasks, such as the large-scale data acquisition and analysis in real-time traffic monitoring\footnote{An example is the crowdsourcing-based traffic and navigation app ``Waze'' (\url{https://www.waze.com}).} or weather monitoring and forecasting~\cite{Niforatos2014} at different places. By focusing on the geospatial data, a new paradigm called \emph{spatial crowdsourcing (SC)~}\cite{Kazemi2012} has received increasing attention in the last few years~\cite{Chen2014,Zhao2016,Guo2018}. Typically, there are three entities in the SC system, including an online SC platform, requesters and workers. As a core component of the SC ecosystem, the SC platform is a broker which allows requesters to post tasks and recruits workers to complete them. Each employed worker then stays at or travels to its target task area to collect and transmit the requested data back. Since the relationship between the SC platform and the workers are incentive-driven, we study the interactions between them to understand and develop an effective mechanism to enable sustainable and efficient operations of the SC systems.

Most existing work assumes that there is always reliable communication infrastructure and enough energy available for workers to complete the data transmission. However, this assumption may not be realistic, especially when the workers have to perform tasks in remote areas without a wireless base station. Moreover, workers can be battery-powered wireless mobile devices. Their energy constraint limits the working time and ultimately affects the task completion. Fortunately, some studies~\cite{Peng2010,Zheng2017,Li2018} in wireless powered sensor networks have illustrated the feasibility of using wireless power transfer (WPT)~\cite{Bi2015} in sensing data collection to prolong the lifetime of sensors. Given this, we consider a paradigm called\emph{ wireless powered spatial crowdsourcing} where the SC platform deploys a mobile base station (BS), e.g., robots, drones or vehicles, to assist the data collection. The mobile BS serves as the infrastructure for communication and wireless power transfer. A typical scenario suitable for applying this paradigm is the information collection in an emergency rescue mission. The requester can be the relief headquarter which needs the SC platform to organize workers to continually transmit the live video or environmental monitoring data from the target task area, e.g., seismic site. These data and data analytics results will significantly help to increase the efficiency of succour. Meanwhile, those workers with battery-powered devices will need wireless charging due to the possible power outage.

To ensure successful and stable operations of the crowdsourcing system, designing an incentive mechanism that stimulates workers' participation and efficiently allocates tasks is essential. A number of studies have proposed mechanisms satisfying various requirements, such as profitability, strategyproofness, i.e., truthfulness, and individual rationality~\cite{Zhao2014,Yang2016}. Nevertheless, in wireless powered spatial crowdsourcing networks, the reward offered by the SC platform to workers can be the wireless power supply, which is location-dependent. The major difference from those existing mechanisms, the incentive of which is based on the monetary reward\footnote{The monetary reward can be tokens, virtual money, reputation, etc.}. The difference introduces a few major issues for incentive mechanism design in wireless powered crowdsourcing networks, and the following questions have to be answered. First, what is the optimal total charging power supply for the SC platform to configure for maximizing its utility? The SC platform can encourage workers to transmit sensed data at a higher transmission rate, i.e., more collected data per unit time, but it is at the cost of a higher power supply. Second, how to allocate the tasks and charging power to workers which are spatially distributed in the target task area? The allocation is based on not only each worker's sensing cost but also the working location, which affects the communication cost and transferred power. Note that the workers' sensing cost and working location can be private information and unknown to the SC platform. Lastly, how to deploy the mobile BS taking the workers' strategic behaviours into account? Since the workers' working locations are private, workers need to report their locations before the mobile BS chooses the best location to deploy.

\begin{figure}[tbh]
\begin{centering}
\includegraphics[width=0.7\columnwidth]{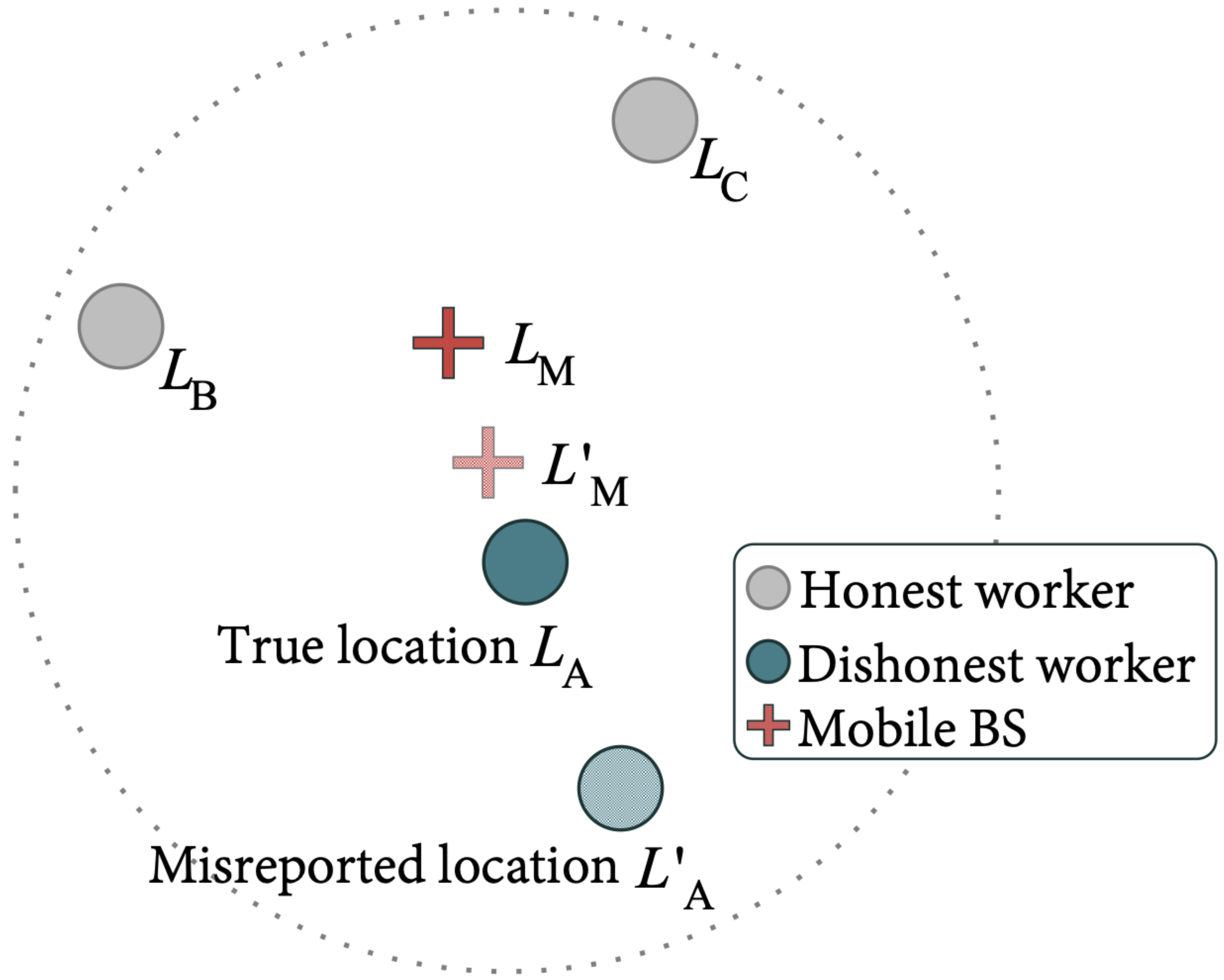}
\par\end{centering}
\caption{An example where a dishonest worker misreports its true location.\label{fig:illustrated-example}}
\end{figure}

Under the assumption of rationality, a worker may dishonestly misreport its location to increase its utility while reducing the SC platform's utility. Figure~\ref{fig:illustrated-example} shows such an example. In the task area, there are one dishonest worker at location $L_{\mathrm{A}}$ and two honest workers respectively at locations $L_{\mathrm{B}}$ and $L_{\mathrm{C}}$. The SC platform would place the mobile BS at $L_{\mathrm{M}}$ for optimal utility if all the workers report true locations $L_{\mathrm{A}},\,\,L_{\mathrm{B}}$ and $L_{C}$. However, the dishonest worker has the incentive to report a fake location $L'_{\mathrm{A}}$, so that according to the reported locations $L'_{\mathrm{A}},\,\,L_{\mathrm{B}}$ and $L_{\mathrm{C}}$, the mobile BS will be deployed at $L'_{\mathrm{M}}$. In this case, the dishonest worker at $L_{\mathrm{A}}$ can be closer to the mobile BS and then enjoy more transferred power from the mobile BS while consuming less power to transmit its sensed data. This inevitably increases other workers' and SC platform's energy consumption and damages their utility. Most current studies on incentive mechanisms for the crowdsourcing system have not addressed such issue yet.
In this paper, we propose a strategyproof and energy-efficient SC framework which jointly solves the problems of task and wireless charging power allocation as well as the truthful working location reporting. In the framework, there are two phases: task allocation phase and data crowdsourcing phase. In the task allocation phase, the SC platform determines and announces a fixed total charging power supply. Each worker interested in participating needs to choose and submits the preferred crowdsourcing plan, i.e., its data transmission rate to the SC platform. In return, they can obtain the corresponding portion of the supplied charging power from the SC platform. We use the Stackelberg game to model the interactions between workers and the SC platform, in which each worker's transmission rate and allocated power can be determined. In the data crowdsourcing phase, the mobile BS requests for workers' working locations. Based on the Moulin's generalization median rule~\cite{Moulin1980}, we present three strategyproof mobile BS deployment mechanisms for the mobile BS to determine its service location. The first one is the classical median mechanism. The other two mechanisms are designed from the Bayesian viewpoint. One is a conventional mechanism which assumes that each worker's working location follows a priori known distribution. For more general scenarios with only historical working location data available, we resort to the advanced deep learning technique to develop another mechanism for higher robustness and larger utility. 
\begin{figure*}[tbh]
	\begin{centering}
	\includegraphics[width=0.75\textwidth]{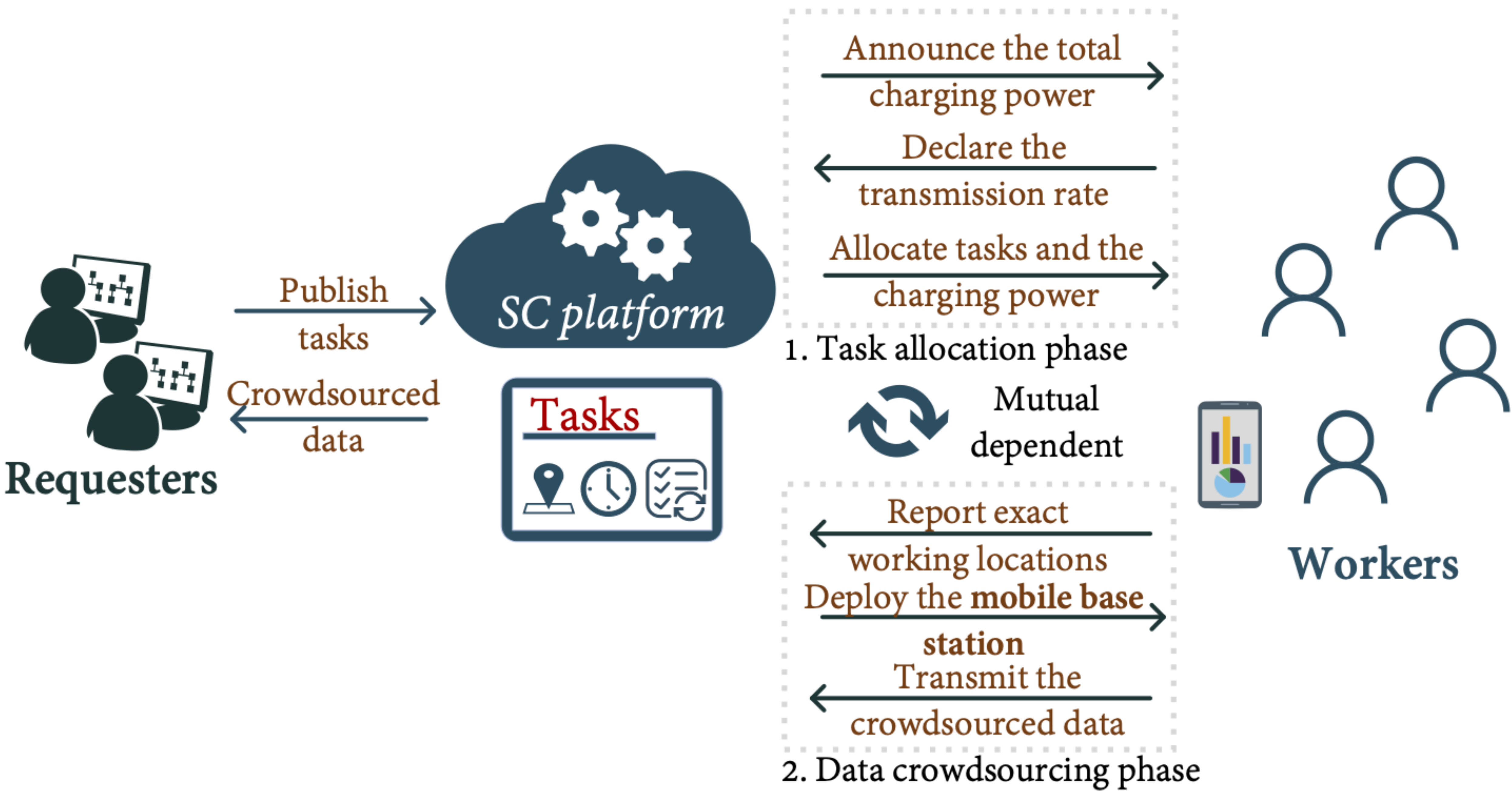}
	\par\end{centering}
	\caption{Wireless powered spatial crowdsourcing system with two phases.\label{fig: SC-system}}
\end{figure*}

The major contributions of this paper can be summarized as follows: 
\begin{itemize} 
	\item We propose a strategyproof and energy-efficient framework for implementing the wireless powered spatial crowdsourcing. The task allocation phase and the data crowdsourcing phase jointly coordinate the task/power allocation and the mobile BS deployment to maximize the SC platform's utility. 
	\item We propose an incentive mechanism for the task and wireless power transfer allocation based on the Stackelberg game model in the task allocation phase. We prove that a unique Nash equilibrium exists among workers' strategies, i.e., the data transmission rates, and the Stackelberg equilibrium can be efficiently calculated to optimize the SC platform's utility. 
	\item In the data crowdsourcing phase, we first present two strategyproof mobile BS deployment mechanisms to prevent the dishonest worker's manipulation while maximizing the SC platform's utility under different scenarios respectively with 1) no prior information 2) prior location distribution. Moreover, for the complex scenario with only historical working location data available, we utilize the deep learning technique and construct a novel deep neural network to design a strategyproof deployment mechanism.
	\item Based on synthetic and real-world datasets, the experimental results illustrate the effectiveness of the proposed incentive mechanisms in assisting the SC platform in allocating the task and charging power efficiently.  In particular, the deep learning based mechanism shows significant improvement in performance and stability compared with the conventional mechanism. 
\end{itemize} 

To the best of our knowledge, this is the first work that investigates the incentive mechanism design in wireless powered spatial crowdsourcing and, for the first time, the deep learning method to address the problem of potential working location misreporting in spatial crowdsourcing systems. 

The rest of the paper is organized as follows. In Section~\ref{sec:Related-Works}, we discuss the related work and motivations in detail. In Section~\ref{sec:System-Model:-Spatial}, we describe the system model of wireless powered spatial crowdsourcing. Section~\ref{sec:Optimal-charging-power} proposes the task and charging power allocation mechanism. In Section~\ref{sec:Bayesian-location-mechanism}, we present three mechanisms for strategyproof mobile BS deployment in the data crowdsourcing phase. In Section~\ref{sec:Experimental-and-simulation}, we provide the experimental results. Finally, we conclude the paper in Section~\ref{sec:Conclusion}. 

\section{Related Work and Motivations\label{sec:Related-Works}}

There have already been studies about the incentive mechanisms in crowdsourcing systems~\cite{Yang2016,Zhang2012}. The authors in~\cite{Yang2016} proposed platform-centric and user-centric incentive mechanisms, respectively based on the Stackelberg game and the reverse auction. Each worker is free to determine its own strategy, i.e., working time or cost, for a reward. Some desirable economic properties, such as strategyproofness\footnote{We use ``strategyproofness'' and ``truthfulness'' interchangeably in this paper.} and individual rationality, are guaranteed in the auction. In~\cite{Zhang2012}, the authors designed a reputation-based incentive mechanism and used the repeated gift-giving game in analyzing the interaction between task requesters and workers. 
 The authors in~\cite{Cardone2013} made use of the historical data about the workers' visiting records to the task locations to investigate workers' skills. For crowdsourcing in wireless-powered task-oriented networks, a game-based distributive incentive mechanism was proposed in~\cite{Yao2018} for reducing energy consumption while ensuring task completion. Particularly, in~\cite{Yao2018}, the authors also used the energy as the reward and introduced an energy bank as the trusted medium of the energy service exchange to avoid using the unreliable and unspecific monetary reward among the workers. However, to the best of our knowledge, most of the existing studies on the crowdsourcing rely on the monetary transfer, i.e., payment, to guarantee the property of truthfulness in reporting private valuations. Moreover, none of the existing work has addressed the issue that a dishonest worker could possibly misreport its working location and manipulate the crowdsourcing system in the data crowdsourcing phase, which cannot be solved using monetary transfer. The study of approximate mechanism design without money was initialized in~\cite{Procaccia2013}, where the authors discussed the strategyproof single facility deployment mechanism in one-dimensional space. The authors in~\cite{Golowich2018} designed two neural network structures, including MoulinNet and RegretNet, to solve the strategyproof multiple facilities location problem in one-dimensional space. Inspired by these works, we propose mobile BS deployment mechanisms for the SC system, which can achieve high utility while guaranteeing the strategyproofness without any money or reward transfer.
\begin{figure*}[tbh]
	\begin{centering}
	\includegraphics[width=0.75\textwidth]{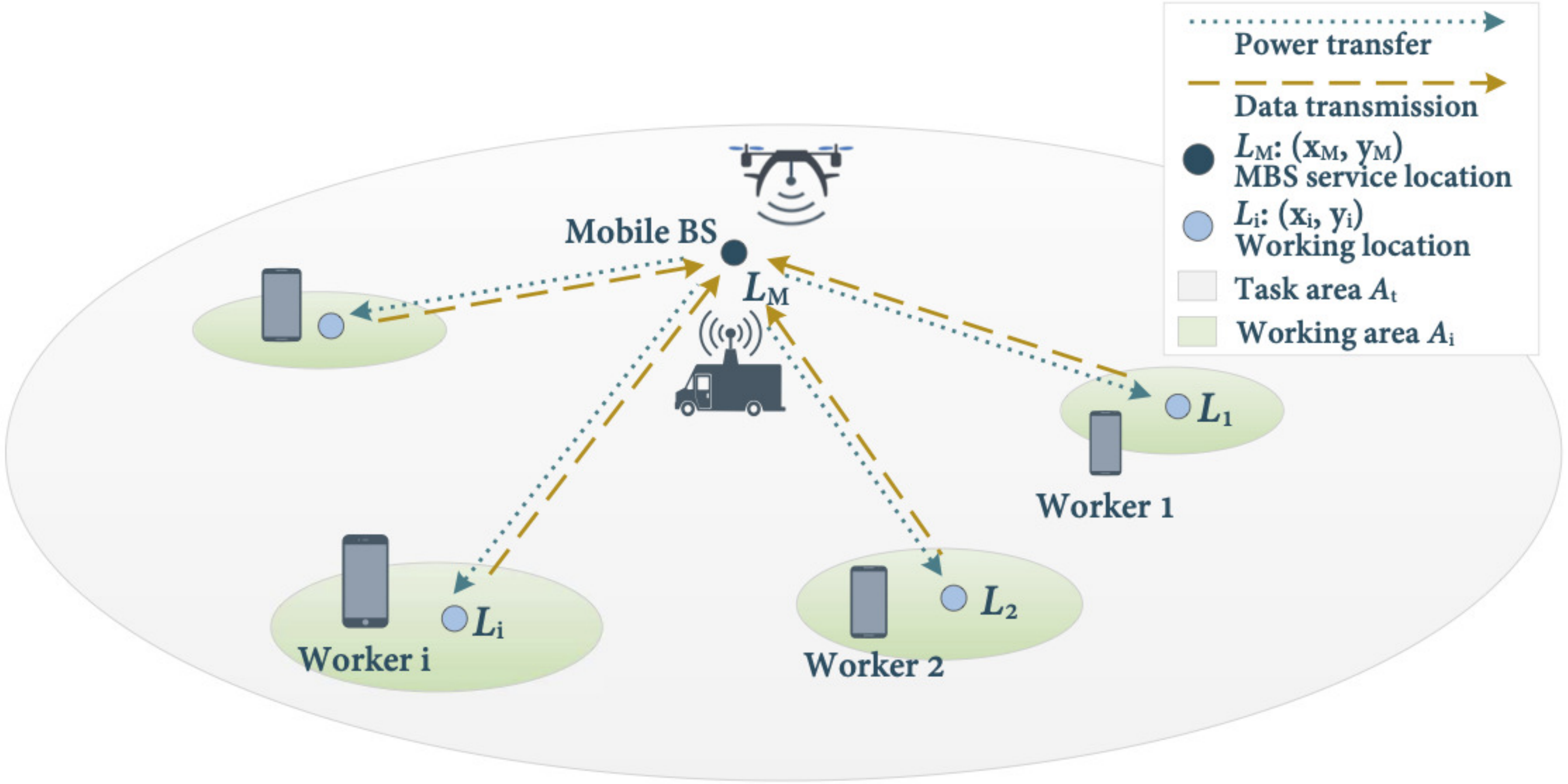}
	\par\end{centering}
	\caption{Data transmission and power transfer in the data crowdsourcing phase.\label{fig:Data-crowdsourcing-stage.}}
	\end{figure*}
\section{System Model: Wireless Powered Spatial Crowdsourcing Market\label{sec:System-Model:-Spatial}}
Figure~\ref{fig: SC-system} depicts the wireless powered spatial crowdsourcing system model where there are three entities, including the requesters, the SC platform residing in the cloud and the workers with mobile sensing devices. The workers can be human, unmanned vehicles or robots. Initially, the requesters publish spatial tasks with requirements, such as the target task area, the task duration, and the sensed data type. Then, the SC platform advertises the task information to\emph{ }workers on behalf of the requesters and collects the crowdsourced or sensed data. As shown in Fig.~\ref{fig:Data-crowdsourcing-stage.}, we denote by $\mathcal{N}=\{1,\ldots,N\}$ the set of workers and denote by $\boldsymbol{A}_{t}$ the task area on a Cartesian coordinate plane. The worker $i$'s working location $L_{i}$ is described by a 2-tuple, i.e., $L_{i\in\mathcal{N}}=(x_{i},y_{i})$. We use $L_{\mathrm{M}}=(x_{\mathrm{M}},y_{\mathrm{M}})\in\boldsymbol{A}_{t}\subseteq\mathbb{R}^{2}$ to represent the deployed mobile BS's service location projected on the XY-plane and use $h$ to denote its height. We assume that each worker knows its preferred area to work, i.e., working area, such as the area near to its commuting route or around home~\cite{Tuncay2012}. In the task area, worker~$i$ has its own working area $\boldsymbol{A}_{i}$ and its working location $L_{i}$ falls in this area, i.e., $L_{i}\in\boldsymbol{A}_{i}\subseteq\boldsymbol{A}_{t}\subseteq\mathbb{R}^{2}$. In this section, we first model the power cost of communication and sensing for the mobile BS and workers in the data crowdsourcing phase. Then, we elaborate on both the task allocation phase and the data crowdsourcing phase and present the problem formulations.

\subsection{Power cost model}

\subsubsection{Worker's power cost}

We consider a frequency division duplexing (FDD) system where sufficient
channels are available to ensure interference-free transmission. Note
that with this assumption, we can better focus on the incentive mechanism
design between the SC platform and workers. Furthermore, we assume
that the communication channels are dominated by line-of-sight (LoS)
links. Given the mobile BS's service location $L_{\mathrm{M}}$, we
can write the worker $i$'s transmission rate according to Shannon's
formula as follows:
\begin{align}
r_{i} & =B\mathrm{\mathrm{\log_{2}}}\left(1+\frac{P_{i}^{\mathrm{t}}\delta}{\sigma^{2}d_{i}^{\alpha}}\right)=B\mathrm{\log_{2}}\left(1+\frac{P_{i}^{\mathrm{t}}g}{d_{i}^{\alpha}}\right)\label{eq:transmission_rate}
\end{align}
where $g=\frac{\delta}{\sigma^{2}}$ is the channel gain to noise ratio (CNR), $\delta$ represents the corresponding channel power gain at the reference distance of $1$ meter, $\sigma^{2}$ is the noise power at the receiver mobile BS, $B$ is the channel bandwidth, $P_{i}^{\mathrm{t}}$ is worker $i$'s data transmission power, and $\alpha\geq2$ is the path-loss exponent. In addition, we define 
\begin{align}
d_{i} & =d_{i}(L_{\mathrm{M}})=d((x_{i},y_{i}),(x_{\mathrm{M}},y_{\mathrm{M}}))\nonumber \\
 & =\sqrt{(x_{i}-x_{\mathrm{M}})^{2}+(y_{i}-y_{\mathrm{M}})^{2}+h^{2}}\label{eq:euclidean_distance}
\end{align}
as the Euclidean distance between the worker $i$ and the mobile BS. Again, $h$ is the height of the mobile BS. Hereby, we can derive the worker $i$'s transmission power as
\begin{equation}
P_{i}^{\mathrm{t}}=\frac{(2^{\frac{r_{i}}{B}}-1)}{g}d_{i}^{\alpha}.\label{eq:transmission_power}
\end{equation}
Besides the power used to transmit data, for the worker $i$, we have the power cost function of data sensing $P_{i}^{\mathrm{s}}=b_{i}r_{i}$ where $b_{i}$ is the energy cost per bit. Here, the power cost of data sensing is linear to the sampling rate~\cite{Dieter2005}, i.e., the transmission rate. Therefore, the worker $i$'s total power cost $P_{i}$ can be expressed as follows:
\begin{align}
P_{i} & =P_{i}^{\mathrm{t}}+P_{i}^{\mathrm{s}}=\frac{(2^{\frac{r_{i}}{B}}-1)}{g}d_{i}^{\alpha}+b_{i}r_{i}.\label{eq:worker_i_total_cost}
\end{align}

\subsubsection{Power cost of the mobile base station}

The mobile BS consumes energy mainly for WPT to workers. If the charging power transferred to the worker $i$ is $P_{i}^{\mathrm{c}}$, the mobile BS at the service location has to consume power $P_{i}^{\mathrm{c}'}$ as follows~\cite{Zhou2013}:
\begin{align}
P_{i}^{\mathrm{c}'} & =\frac{P_{i}^{\mathrm{c}}d_{i}^{\alpha}}{\eta\Gamma}=P_{i}^{\mathrm{c}}d_{i}^{\alpha}\kappa,\label{eq:charging_power}
\end{align}
where $\kappa=\frac{1}{\eta\Gamma}$, $0<\eta<1$ denotes the receiver energy conversion efficiency, $\Gamma$ denotes the combined antenna gain at the reference distance of $1$ meter.

\subsection{Utility function in the wireless powered spatial crowdsourcing\textmd{\normalsize{}}system}

We define the utility of the crowdsourced data based on the transmission rate, which combines two common metrics, i.e., the data size and timeliness. For example, the requesters may perform the data analysis and prediction based on the real-time crowdsourced data. Higher data transmission rate means that the requesters can process more data during a unit time and yield more accurate prediction results. The utility of the crowdsourced data is equivalent to the utility of the SC task completion. The utility $q$ of data collected from the SC task completion is calculated by
\begin{align}
q(\mathbf{r}) & =a_{1}\log(1+\sum_{i\in\mathcal{N}}\log(1+a_{2}r_{i})),\label{eq:sensing_quality}
\end{align}
where $\mathbf{r}=(r_{1},r_{2},\ldots,r_{N})$ is the transmission rate vector reported by workers, $a_{1}$ and $a_{2}$ are parameters. The inner logarithmic function reflects the SC platform\textquoteright s diminishing return of the worker $i$'s contribution, and the outer logarithmic function reflects the diminishing return of all participating workers' contributions~\cite{Jiao2018,Yang2012}. In this paper, the mobile BS serves as a dedicated power transmitter which applies the directional beamforming technique~\cite{Huang2014}. Taking the power cost of WPT~(\ref{eq:charging_power}) into consideration, the SC platform's utility function can be expressed as~\cite{Huang2014}
\begin{align}
u_{m} & =q(\mathbf{r})-\sum_{i\in\mathcal{N}}P_{i}^{\mathrm{c}'}\nonumber \\
 & =a_{1}\log(1+\sum_{i\in\mathcal{N}}\log(1+a_{2}r_{i}))-\sum_{i\in\mathcal{N}}P_{i}^{\mathrm{c}}d_{i}^{\alpha}\kappa.\label{eq:SC_platform_utility_func}
\end{align}
Similarly, we obtain the worker $i$'s utility function as
\begin{align}
u_{i} & =P_{i}^{\mathrm{c}}-P_{i}=P_{i}^{\mathrm{c}}-\frac{(2^{\frac{r_{i}}{B}}-1)}{g}d_{i}^{\alpha}-b_{i}r_{i}.\label{eq:worker_i_utility_func}
\end{align}

\subsection{The procedure of wireless powered spatial crowdsourcing}

Note that we aim to maximize the SC platform's utility. Recalling the utility functions in~(\ref{eq:SC_platform_utility_func}) and (\ref{eq:worker_i_utility_func}), how to determine each worker's transmission rate and charging power as the reward and where to deploy the mobile BS are two critical issues for utility maximization. 

\subsubsection{Task allocation phase\label{subsec:Charging-Power-Allocation}}

Before the mobile BS departs to collect data and workers execute the assigned tasks, the SC platform announces a total charging power supply $P_{\mathrm{c}}$ ($P_{\mathrm{c}}=\sum_{j\in\mathcal{N}}P_{j}^{\mathrm{c}}$) to assist workers in the data crowdsourcing. The charging power $P_{i}^{\mathrm{c}}$ transferred to worker $i$ is proportional to its contribution (the data transmission rate), i.e., $P_{i}^{\mathrm{c}}=\frac{r_{i}}{R}P_{\mathrm{c}}=\frac{r_{i}}{\sum_{j\in\mathcal{N}}r_{j}}P_{\mathrm{c}}$. Based on the sensing tasks and the other workers' responses, each worker reports the preferred data rate $r_{i}$ to maximize its own utility. In practice, the SC platform may serve as a relay to receive and broadcast the workers' responses. As workers have not determined the suitable working place and perform the allocated task, they are exposed to uncertainty of working location $L_{i}$ and the mobile BS's service location $L_{\mathrm{M}}$ which are only known in the next data crowdsourcing phase. We assume that the workers are \emph{risk-averse}, which means that they choose to minimize the uncertainty and avoid any possible loss in the future. This concept can be found in the well-known prospect theory~\cite{Kahneman2013}. A common example is that a majority of people prefer to deposit money at the bank for safekeeping and low return instead of buying financial products with a high risk of loss. Note that given the power supply and other workers' transmission rates, the worker $i$'s utility function in~(\ref{eq:worker_i_utility_func}) is monotonically decreasing with $d_{i}$. Since the worker $i$ knows its working area $\boldsymbol{A}_{i}$ and the task area $\boldsymbol{A}_{t}$, it can obtain the maximum value of $d_{i}$, i.e., $D_{i}=\max_{L_{M}\in\boldsymbol{A}_{t},L_{i}\in\boldsymbol{A}_{i}}d_{i}$. Therefore, if the worker $i$ plans the transmission rate $r_{i}$ for the worst case where $D_{i}$ is its distance from the mobile BS, the worker $i$ will achieve the utility which is not lower than the worst case in the data crowdsourcing phase. In addition, we use $\mathbf{r}_{-i}=(r_{1},\ldots,r_{i-1},r_{i+1},\ldots,r_{N})$ to denote the reported transmission rate vector for all workers except the worker $i$. Hereby, the worker $i$'s utility function in the task allocation phase can be expressed as
\begin{equation}
\bar{u}{}_{i}(r_{i},\mathbf{r}_{-i},P_{\mathrm{c}})=\frac{r_{i}}{\sum_{j\in\mathcal{N}}r_{j}}P_{\mathrm{c}}-\frac{(2^{\frac{r_{i}}{B}}-1)}{g}D_{i}^{\alpha}-b_{i}r_{i}.\label{eq:worker_i_utility_task_allocation_stage}
\end{equation}
The SC platform's utility in~(\ref{eq:SC_platform_utility_func}) is rewritten as
\begin{align}
\bar{u}{}_{m} & (P_{\mathrm{c}},\mathbf{r})=a_{1}\log(1+\sum_{i\in\mathcal{N}}\log(1+a_{2}r_{i}))\nonumber \\
 & \qquad\qquad-\sum_{i\in\mathcal{N}}\frac{r_{i}}{\sum_{j\in\mathcal{N}}r_{j}}P_{\mathrm{c}}D_{i}^{\alpha}\kappa.\label{eq:SC_platform_utility_task_allocation_phase}
\end{align}

\subsubsection{Data crowdsourcing phase\label{subsec:Data-crowdsoucing-phase}}

In the task allocation phase, the total charging power supply $P_{\mathrm{c}}$, each worker's allocated charging power $P_{i}^{\mathrm{c}}$ and transmission rate $r_{i}$ have been determined. Each worker decides the working location according to the task and its available working area. For example, if the task requires collecting data about road traffic condition, workers may choose the roadside or crossing. As our paper mainly focuses on establishing a spatial crowdsourcing market with wireless energy transfer and designing relevant trading mechanisms, how to choose a good working location is beyond our scope. Once working locations are decided, they will travel to the working locations and the SC platform sends out the mobile BS to serve the workers. However, the mobile BS has to know each worker's working location. Then, it can determine the service location $L_{M}$ for maximizing the SC platform's utility. The worker $i$'s and the SC platform's utility functions in the data crowdsourcing phase can be respectively expressed as
\begin{equation}
\hat{u}{}_{i}(L_{\mathrm{M}})=\frac{r_{i}}{\sum_{j\in\mathcal{N}}r_{j}}P_{c}-\frac{(2^{\frac{r_{i}}{B}}-1)}{g}d_{i}^{\alpha}(L_{i},L_{\mathrm{M}})-b_{i}r_{i}\label{eq:worker_i_utility_dc_phase}
\end{equation}
 and 
\begin{align}
\hat{u}{}_{m} & (L_{\mathrm{M}})=a_{1}\log(1+\sum_{i\in\mathcal{N}}\log(1+a_{2}r_{i}))\nonumber \\
 & \qquad\qquad-\sum_{i\in\mathcal{N}}\frac{r_{i}}{\sum_{j\in\mathcal{N}}r_{j}}P_{\mathrm{c}}d_{i}^{\alpha}(L_{i},L_{\mathrm{M}})\kappa.\label{eq:SC_platform_utility_dc_phase}
\end{align}
To make workers reveal their private working location $L_{i}$, the mobile BS organizes the following voting process on the spot. 
\begin{enumerate}
\item The mobile BS first broadcasts its deployment mechanism, i.e, the
mechanism or rule to place the mobile BS according to the locations
reported by workers, to the task area. 
\item Once receiving the notification about the deployment mechanism, each
worker sends its working location $L_{i}$ to the mobile BS.
\item Based on the collected locations and the deployment mechanism, the
service location $L_{\mathrm{M}}$ is calculated for the mobile BS
to deploy.
\end{enumerate}
Let $\mathsf{M}$ denote the applied deployment mechanism which takes the workers' reported working location vector $\mathbf{L}=(L_{1},\ldots,L_{i},\ldots,L_{\mathrm{N}})$ as input and outputs the mobile BS's service location $L_{\mathrm{M}}$, i.e., $L_{\mathrm{M}}=\mathsf{M}(\mathbf{L})$. During the above voting process, a worker $i$ may have an incentive to improve its own utility in~(\ref{eq:worker_i_utility_dc_phase}) by misreporting its true working location $L_{i}$. To make the location voting process robust and implementable, our designed mobile BS deployment mechanism should have the property of strategyproofness (truthfulness), which is defined as follows: 
\begin{defn}
	(Strategyproofness) Regardless of other workers' reported locations, a worker $i$ cannot increase the utility by misreporting its working location $L_{i}$. Formally, given a deployment mechanism $\mathsf{M}$ and a misreported location $L'_{i}$, we have 
\begin{equation}
\hat{u}_{i}(\mathsf{M}((L_{i},\mathbf{L}_{-i})))\geq\hat{u}_{i}(\mathsf{M}((L'_{i},\mathbf{L}_{-i})))\:\forall L_{i}'\neq L_{i}
\end{equation}
where $\mathbf{L}_{-i}$ is the vector containing all workers' working
locations except the worker $i$'s.
\end{defn}

\subsubsection{Mutual Dependence}

The task allocation phase and the data crowdsourcing phase are mutually dependent. On the one hand, each worker's transmission rate in data crowdsourcing is determined from the task allocation phase. On the other hand, a prerequisite of the successful charging power allocation is to guarantee that the data crowdsourcing phase cannot be strategically manipulated. The untruthful or dishonest worker may overestimate its risk preference, i.e., the maximum distance $D_{i}$, due to its deliberate manipulation. Both the two phases affect the efficient use of the power as well as all the participants' utilities. 

\section{Task and Wireless Transferred Power Allocation Mechanism\label{sec:Optimal-charging-power}}

We utilize the Stackelberg game approach~\cite{Fudenberg1991} to analyze the model introduced in the task allocation phase (Section~\ref{subsec:Charging-Power-Allocation}). There are two levels in the Stackelberg game. In the first (upper) level, the SC platform acts as the leader which strategizes and announces the total charging power supply $P_{\mathrm{c}}$. In the second (lower) level, each worker is the follower which determines the strategy, i.e., the preferred transmission rate $r$, to maximize its utility. Mathematically, the SC platform chooses the strategy $P_{\mathrm{c}}$ by solving the following optimization problem:
\[
(\mathrm{P1})\;\max_{P_{c}\geq0}\bar{u}{}_{m}(P_{\mathrm{c}},\mathbf{r}).
\]
Meanwhile, the worker $i$ makes the decision on its reported $r_{i}$
to solve the following problem:
\[
(\mathrm{P2})\;\max_{r_{i}\geq0}\bar{u}{}_{i}(r_{i},\mathbf{r}_{-i},P_{\mathrm{c}}).
\]
The objective of the Stackelberg game is to find the \emph{Stackelberg
Equilibrium (SE)}. We next introduce the concept of the SE for our
proposed model.
\begin{defn}
	(Stackelberg Equilibrium) Let $\tilde{P}_{c}$ be a solution for Problem $\mathrm{P1}$ and $\tilde{\mathbf{r}}$ be a solution for Problem $\mathrm{P2}$ of the workers. Then, a point $(\tilde{P}_{\mathrm{c}},\tilde{\mathbf{r}})$ is the SE for the proposed Stackelberg game if it satisfies the following conditions:
\begin{equation}
\bar{u}{}_{m}(\tilde{P}_{\mathrm{c}},\tilde{\mathbf{r}})\geq\bar{u}{}_{m}(P_{\mathrm{c}},\tilde{\mathbf{r}}),
\end{equation}
\begin{equation}
\bar{u}{}_{i}(\tilde{r}_{i},\tilde{\mathbf{r}}_{-i},\tilde{P}_{\mathrm{c}})\geq\bar{u}{}_{i}(r_{i},\tilde{\mathbf{r}}_{-i},\tilde{P}_{\mathrm{c}}),
\end{equation}
for any $(P_{\mathrm{c}},\mathbf{r})$ with $P_{c}\geq0$ and $\mathbf{r}\succeq0$.
\end{defn}
In general, the first step to obtain the SE is to find the perfect \emph{Nash Equilibrium (NE)~\cite{Fudenberg1991}} for the non-cooperative transmission Rate Determination Game (RDG) in the lower level. Then, we can optimize the strategy of the SC platform at the upper level. Given a fixed $P_{c}$, the NE is defined as a set of strategies $\mathbf{r}^{\mathrm{ne}}=(r_{1}^{\mathrm{ne}},\ldots,r_{N}^{\mathrm{ne}})$ that no worker can improve utility by unilaterally changing its own strategy while other workers' strategies are kept unchanged. Since workers are rational and not willing to provide service for a negative utility, they shall set $r_{i}=0$ if $\bar{u}{}_{i}(r_{i},\mathbf{r}_{-i},P_{\mathrm{c}})\leq0$. To analyze the NE, we introduce the concept of the \emph{concave game} and the theorem about the \emph{existence} and \emph{uniqueness} of NE in the concave game. 
\begin{defn}
	(Concave game~\cite{Rosen1965})\emph{ \label{def:(Concave-game)}}A game is called concave if each worker $i$ chooses a strategy $r_{i}$ to maximize utility $\bar{u}{}_{i}$, where $\bar{u}{}_{i}$ is concave in $r_{i}$. 
\end{defn}
\begin{thm}
	(\cite{Rosen1965}) \label{thm:NE-existance-uniqueness-Concave-games}Concave games have (possibly multiple) Nash Equilibrium. Define $N\times N$ matrix function $\mathbf{H}$ in which $\mathbf{H}_{ij}=\frac{\partial^{2}\bar{u}{}_{i}}{\partial r_{i}\partial r_{j}}$,$i,j\in\mathcal{N}$. Let \textup{$\mathbf{H}^{\mathrm{T}}$ denote the transpose of $\mathbf{H}$.} If $\mathbf{H}+\mathbf{H}^{\mathrm{T}}$ is strictly negative definite, then the Nash equilibrium is unique.
\end{thm}
Hereby, we calculate the first-order and second-order derivatives of the worker $i$'s utility function $\bar{u}{}_{i}(r_{i},\mathbf{r}_{-i},P_{c})$ with respect to $r_{i}$ as follows: 
\begin{equation}
\frac{\partial\bar{u}{}_{i}}{\partial r_{i}}=\frac{P_{\mathrm{c}}\sum_{k\in\mathcal{N}_{-i}}r_{k}}{(\sum_{j\in\mathcal{N}}r_{j})^{2}}-\frac{D_{i}^{\alpha}\ln2}{B}2^{\frac{r_{i}}{B}}-b_{i},\label{eq:first-derivative}
\end{equation}
\begin{equation}
\frac{\partial^{2}\bar{u}{}_{i}}{\partial r_{i}^{2}}=-\frac{2P_{\mathrm{c}}\sum_{k\in\mathcal{N}_{-i}}r_{k}}{(\sum_{k\in\mathcal{N}}r_{k})^{3}}-\frac{D_{i}^{\alpha}\ln^{2}2}{B^{2}}2^{\frac{r_{i}}{B}}.\label{eq:second-derivative}
\end{equation}
Since $\frac{\partial^{2}\bar{u}{}_{i}}{\partial r_{i}^{2}}<0$, $\bar{u}{}_{i}(r_{i},\mathbf{r}_{-i},P_{c})$ is a strictly concave function with respect to $r_{i}$. Then, the non-cooperative RDG is a concave game and the NE exists when $\sum_{j\in\mathcal{N}_{-i}}r_{j}>0$. Otherwise the worker $i$'s best strategy does not exist. Given any $P_{c}>0$ and any strategy profile $\mathbf{r}_{-i}$ $(\sum_{j\in\mathcal{N}_{-i}}r_{j}>0)$, the worker $i$'s best response strategy $\gamma_{i}$ exists and is unique. To prove the uniqueness of the NE, we also calculate the second-order mixed partial derivative of $\bar{u}{}_{i}$ for $i\in\mathcal{N}$ with respect to $r_{j\in\mathcal{N}_{-i}}$ as follows: 
\[
\frac{\partial^{2}\bar{u}{}_{i}}{\partial r_{j}^{2}}=\frac{2r_{i}}{(\sum_{j\in\mathcal{N}}r_{j})^{3}}P_{\mathrm{c}},\,\,\frac{\partial^{2}\bar{u}{}_{i}}{\partial r_{i}\partial r_{j}}=\frac{r_{i}-\sum_{k\in\mathcal{N}_{-i}}r_{k}}{(\sum_{k\in\mathcal{N}}r_{k})^{3}}P_{\mathrm{c}},
\]
where $\frac{\partial^{2}\bar{u}{}_{i}}{\partial r_{j}^{2}}\geq0$ and $\frac{\partial^{2}\bar{u}{}_{i}}{\partial r_{i}\partial r_{j}}\leq0$ if $r_{i}\leq\sum_{k\in\mathcal{N}_{-i}}r_{k},\,\forall i\in\mathcal{N}$. Then, we have the specific expression of the matrix function $\mathbf{H}$ defined in Theorem~\ref{thm:NE-existance-uniqueness-Concave-games}. Furthermore, the matrix function $\mathbf{H}+\mathbf{H}^{\mathrm{T}}$ can be decomposed into a sum of several $N\times N$ matrix functions:
$\mathbf{H}+\mathbf{H}^{\mathrm{T}}=\mathbf{U}+\mathbf{V}+\sum_{k\in\mathcal{N}}$$\mathbf{C}^{k}$,
where $\mathbf{U}_{ij}=\begin{cases}
0 & i\neq j\\
\frac{\partial^{2}\bar{u}{}_{i}}{\partial R_{i}^{2}} & i=j
\end{cases}$, $\mathbf{V}_{ij}=\sum_{k\in\mathcal{N}}\frac{\partial^{2}\bar{u}{}_{k}}{\partial r_{i}\partial r_{j}}$
and $\mathbf{C}_{ij}^{k}=$$\begin{cases}
0 & i=k\,\mathrm{or}\,j=k\\
-\frac{\partial^{2}\bar{u}{}_{k}}{\partial r_{i}\partial r_{j}} & otherwise.
\end{cases}$ Let $\mathbf{I}$ denote the sum of $\mathbf{C}_{ij}^{k}$ over $\mathcal{N}$,
i.e., $\mathbf{I}=\sum_{k\in\mathcal{N}}\mathbf{C}^{k}$. Since $\frac{\partial^{2}\bar{u}{}_{i}}{\partial r_{i}^{2}}<0$
and $\frac{\partial^{2}\bar{u}{}_{i}}{\partial r_{i}\partial r_{j}}\leq0$ , if $r_{i}\leq\sum_{k\in\mathcal{N}_{-i}}r_{k},\,\forall i\in\mathcal{N}$, we can find that $\mathbf{U}$ is strictly negative definite, and $\mathbf{V}$ and $\mathbf{I}$ are negative semi-definite. Thus, $\mathbf{H}+\mathbf{H}^{\mathrm{T}}$ is proved to be strictly negative definite which shows the NE in the RDG is unique. In other words, once the SC platform decides a strategy $P_{\mathrm{c}}$, the workers' strategies, i.e., the transmission rates, will be uniquely determined. We then can use the iterative best response~\cite{Han2012} to find the SE point $\tilde{P}_{\mathrm{c}}$ in the first level, i.e., the optimal strategy of $P_{\mathrm{c}}$.

\section{Mobile BS Deployment Mechanisms in Data Crowdsourcing Phase \label{sec:Bayesian-location-mechanism}}

Given the SE points ($\tilde{P}_{c},\tilde{\mathbf{\mathbf{r}}}$) calculated from the task allocation phase, we use $\hat{\mathcal{N}}=\{1,\ldots,\hat{N}\}$ (break ties randomly) to represent the set of \emph{employed workers} whose transmission rate $\tilde{r}_{i}>0$. Hence, the specific problems for the SC platform in the data crowdsourcing phase is
\begin{align}
\max_{L_{\mathrm{M}}\in\boldsymbol{A}_{t}}\hat{u}_{m}(L_{\mathrm{M}}) & =a_{1}\log(1+\sum_{i\in\mathcal{N}}\log(1+a_{2}\tilde{r}_{i}))\nonumber \\
 & \,\,\,\,\,\,\,-\sum_{i\in\mathcal{\hat{N}}}\frac{\tilde{r}_{i}}{\sum_{j\in\hat{\mathcal{N}}}\tilde{r}_{j}}\tilde{P_{c}}d_{i}^{\alpha}(L_{i},L_{\mathrm{M}})\kappa.\label{eq:SC_utility_data_crowdsourcing}
\end{align}
Based on workers' reported working locations, the SC platform decides the mobile BS's location to maximize its utility. For simplicity, we write the equivalent problems as follows:
\begin{align}
\min_{L_{\mathrm{M}}\in\boldsymbol{A}_{t}}\hat{l}_{m}(L_{\mathrm{M}}) & =\sum_{i\in\hat{\mathcal{N}}}\frac{\tilde{r}_{i}}{\sum_{j\in\mathcal{N}}\tilde{r}_{j}}\tilde{P_{\mathrm{c}}}d_{i}^{\alpha}(L_{i},L_{\mathrm{M}})\kappa,\label{eq:SC_cost_function}
\end{align}
where $\hat{l}_{m}(L_{\mathrm{M}})$ is the \emph{crowdsourcing cost} of SC platform. Minimizing the SC platform's crowdsourcing cost is equivalent to maximizing its utility. Similarly, the worker $i$'s utility and crowdsourcing cost can be respectively expressed as 
\begin{align}
\hat{u}_{i}(L_{\mathrm{M}}) & =\frac{\tilde{r}_{i}}{\sum_{j\in\hat{\mathcal{N}}}\tilde{r}_{j}}\tilde{P_{c}}-\frac{(2^{\frac{\tilde{r}_{i}}{B}}-1)}{g}d_{i}^{\alpha}(L_{i},L_{\mathrm{M}})-b_{i}\tilde{r}_{i},\label{eq:Worker_i_utility_data_crowdsourcing}
\\\hat{l}_{i}(L_{\mathrm{M}}) & =\frac{(2^{\frac{\tilde{r}_{i}}{B}}-1)}{g}d_{i}^{\alpha}(L_{i},L_{\mathrm{M}}).\label{eq:Worker_i_cost_function}
\end{align}

To address the mobile BS's location problem introduced in Section~\ref{subsec:Data-crowdsoucing-phase}, we first present the classical median mechanism and analyze its worst-case performance. Then, we propose a conventional mechanism to improve the utility of the SC platform in expectation. For more general scenarios and achieving better performance, we also propose a deep learning based strategyproof mechanism. The design rationale of the deep neural network is the Moulin's generalized median mechanism.

\subsection{Conventional strategyproof mechanism under Bayesian settings\label{subsec:Conventional-mechanism} }

We first introduce an important concept of \emph{$2$-dimensional single-peaked preference} for the discussed problem.

\begin{defn}
	\emph{\label{def:2-dimensional-single-peaked}($2$-dimensional single-peaked preference~\cite{Barbera1993})} Let $\mathbf{L}_{\mathrm{M}}$ be the set of possible mobile BS's service locations output by the deployment mechanism $\mathsf{M}$ on the XY-plane where $X$ and $Y$ are respectively a one-dimensional axis. The worker $i$\textquoteright s preference for the mobile BS's location is $2$-dimensional single-peaked with respect to $(X,Y)$ if 1) there is a single most-preferred location outcome $\tilde{L}_{i}^{\mathrm{M}}\in\mathbf{L}_{\mathrm{M}}$, and 2) for any two outcomes $L'_{\mathrm{M}},L''_{\mathrm{M}}\in\mathbf{L}_{\mathrm{M}}$, $L'_{\mathrm{M}}\succeq_{i}L''_{\mathrm{M}}$ whenever $L''_{\mathrm{M}}<_{\rho}L'_{\mathrm{M}}<_{\rho}\tilde{L}_{i}^{\mathrm{M}}$ or $\tilde{L}_{i}^{\mathrm{M}}<_{\rho}L'_{\mathrm{M}}<_{\rho}L''_{\mathrm{M}}$ for $\forall\rho\in\{X,Y\}$, i.e., both $X$ and $Y$ axes. 
\end{defn}

In the above definition, $L'_{\mathrm{M}}\succeq_{i}L''_{\mathrm{M}}$ means that $L'_{\mathrm{M}}$ is preferred by worker $i$ to $L''_{\mathrm{M}}$. ``$<_{\rho}$'' is a strict ordering by worker $i$ on the dimension $\rho$. An explanation of this condition is that $L'_{\mathrm{M}}$ is preferred by worker $i$ to $L''_{\mathrm{M}}$ as long as $L'_{\mathrm{M}}$ is nearer to its most-preferred location $\tilde{L}_{i}^{\mathrm{M}}$ on each dimension.

\begin{prop}
	\label{prop:2-dimensional-single-peaked-dcp}In the data crowdsourcing phase, the worker's preference for the mobile BS's service location is $2$-dimensional single-peaked.
\end{prop}
\begin{IEEEproof}
We first expand the worker $i$'s crowdsourcing cost function given in~(\ref{eq:Worker_i_cost_function}) as $\hat{l}_{i}(L_{\mathrm{M}})=\hat{l}_{i}(x_{\mathrm{M}},y_{\mathrm{M}})=\frac{(2^{\frac{\tilde{r}_{i}}{B}}-1)}{g}\left((x_{i}-x_{\mathrm{M}})^{2}+(y_{i}-y_{\mathrm{M}})^{2}+h^{2}\right)^{\frac{\alpha}{2}}$. We can then show that $\hat{l}{}_{i}$ is convex with respect to $(x_{\mathrm{M}},y_{\mathrm{M}})$ and there is a unique optimal solution $\tilde{L}_{i}^{\mathrm{M}}=(x_{i},y_{i})$ to minimizing the cost. In other words, the worker $i$'s most preferred mobile BS's service location is its working location, i.e., $\tilde{L}_{i}^{\mathrm{M}}=L_{i}=(x_{i},y_{i})$, which satisfies the first condition in Definition~\ref{def:2-dimensional-single-peaked}. In the task area $\boldsymbol{A}_{t}$, we randomly choose two locations $L'_{\mathrm{M}}=(x'_{\mathrm{M}},y'_{\mathrm{M}})$, $L''_{\mathrm{M}}=(x''_{\mathrm{M}},y''_{\mathrm{M}})\in\boldsymbol{A}_{t}$. Note that the convexity of $\hat{l}{}_{i}$ guarantees the convexity on one dimension if fixing the variable on the other dimension is fixed. $L''_{\mathrm{M}}<_{X}L'{}_{\mathrm{M}}<_{X}\tilde{L}_{i}^{\mathrm{M}}$ implies that $\hat{l}{}_{i}((x_{i},y))<\hat{l}{}_{i}((x'_{\mathrm{M}},y))<\hat{l}{}_{i}((x''_{\mathrm{M}},y))$ for any $y$ on axis $Y$ and then $\left|x_{i}-x'_{\mathrm{M}}\right|<\left|x_{i}-x''_{\mathrm{M}}\right|$. We can have the similar implication from $L''_{\mathrm{M}}<_{Y}L'{}_{\mathrm{M}}<_{Y}\tilde{L}_{i}^{M}$. If $L''_{\mathrm{M}}<_{X}L'{}_{\mathrm{M}}<_{X}\tilde{L}_{i}^{\mathrm{M}}$ and $L''_{\mathrm{M}}<_{Y}L'{}_{\mathrm{M}}<_{Y}\tilde{L}_{i}^{M}$ are both satisfied, we can have $(x_{i}-x'_{\mathrm{M}})^{2}+(y_{i}-y'_{\mathrm{M}})^{2}<(x_{i}-x''_{\mathrm{M}})^{2}+(y_{i}-y''_{\mathrm{M}})^{2}$ and thus $\hat{l}_{i}(L'_{\mathrm{M}})=\hat{l}{}_{i}((x'_{\mathrm{M}},y'_{\mathrm{M}}))<\hat{l}_{i}(L''_{\mathrm{M}})=\hat{l}{}_{i}((x''_{\mathrm{M}},y''_{\mathrm{M}}))$. Therefore, the worker $i$ prefers $L'_{\mathrm{M}}$ to $L''_{\mathrm{M}}$, i.e., $L'_{\mathrm{M}}\succeq_{i}L''_{\mathrm{M}}$, which proves the condition 2 in Definition~\ref{def:2-dimensional-single-peaked} and completes the proof.
\end{IEEEproof}
\begin{thm}
\label{thm:(Moulin's-one-dimensional-genera)}(Moulin's one-dimensional
generalized median mechanism~\cite{Moulin1980}) A mechanism $\mathsf{M}$
for single-peaked preferences in a one-dimensional space is strategyproof
and anonymous if and only if there exist $\hat{N}+1$ constants $\tau_{1},\ldots,\tau_{\hat{N}+1}\in\mathbb{R}\cup(-\infty,+\infty)$
such that: 
\begin{equation}
\mathsf{M}(\mathbf{L}^{\mathrm{M}})=\mathrm{median}(\tilde{L}_{1}^{\mathrm{M}},\ldots,\tilde{L}_{N}^{\mathrm{M}},\tau_{1},\ldots,\tau_{N+1})
\end{equation}
\emph{where }\textup{\emph{$\mathbf{L}^{\mathrm{M}}=\{\tilde{L}_{1}^{\mathrm{M}},\ldots,\tilde{L}_{\hat{N}}^{\mathrm{M}}\}$
is the set of workers' most-preferred mobile BS's locations and}}\emph{
}\textup{\emph{$\mathrm{median}$ is the median function. }}An outcome
rule $\mathsf{M}$ is anonymous, if for any permutation $\boldsymbol{T}'$
of $\boldsymbol{T}$, we have $\mathsf{M}(\boldsymbol{T}')=\mathsf{M}(\boldsymbol{T})$
for all $\boldsymbol{T}$.
\end{thm}
\begin{thm}
\label{thm:Multi-dimensional-generalized-median}(Multi-dimensional
generalized median mechanism~\cite{Barbera1993}) A mechanism for
multi-dimensional single-peaked preferences in a multi-dimensional
space is strategyproof and anonymous if and only if it is an $m$-dimensional
generalized median mechanism, which straightforwardly applies the
one-dimensional generalized median mechanism on each of the $m$ dimensions.
\end{thm}
\begin{algorithm}[tbh]
	\scriptsize
	\begin{algorithmic}[1]
	\Require{Workers' reported locations $\mathbf{L}=(L_1,\ldots, L_i, \ldots,L_{\hat{N}})$.}
	\Ensure{Mobile BS's service location $L_{\mathrm{M}}=(x_{\mathrm{M}},y_{\mathrm{M}})$.}
	\Begin
		\State{Repectively sort the x coordinates $\mathbf{x}=(x_1,\dots, x_{\hat{N}})$ and y coordinates $\mathbf{y}=(y_1, \dots,y_{\hat{N}})$ of workers' locations in ascending order.}
		\If{$\hat{N}$ is odd}
			\State{$x_{\mathrm{M}} \gets x_{\frac{{\hat{N}}+1}{2}}, y_{\mathrm{M}} \gets y_{\frac{\hat{N}+1}{2}}$}
		\Else
			\State{$x_{\mathrm{M}} \gets \frac{x_{\frac{\hat{N}}{2}}+x_{\frac{\hat{N}}{2}+1}}{2}, y_{\mathrm{M}} \gets \frac{y_{\frac{\hat{N}}{2}}+y_{\frac{\hat{N}}{2}+1}}{2}$}
		\EndIf 
	\End
	\end{algorithmic} 
	\caption{MED mechanism\label{alg:MED-mechanism}}
\end{algorithm}

A straightforward benchmark mechanism is the median mechanism~\cite{Moulin1980,Barbera1993}, as shown in Algorithm~\ref{alg:MED-mechanism}. We simply name it as MED mechanism, i.e., $\mathsf{M}_{\mathrm{MED}}$. This algorithm directly computes the median of workers' reported locations as the mobile BS's service location. Apparently, it is a special case of the multi-dimensional generalized median mechanism, so it is strategyproof. We next analyze its performance by comparing it with the optimal mechanism $\mathsf{M}_{\mathrm{OPT}}$ which achieves the maximum utility of the SC platform without considering incentive constraints. Let $\tilde{r}_{\max}$ and $\tilde{r}_{\min}$ respectively denote the maximum and the minimum transmission rate among workers, i.e., $\tilde{r}_{\max}=\max(\tilde{\mathbf{r}}),\tilde{r}_{\min}=\min(\tilde{\mathbf{r}})$.

\begin{prop}
	\label{prop:The-approximation-ratio}The benchmark MED mechanism $\mathsf{M}_{\mathrm{MED}}$ has an approximation ratio \textup{$2^{\frac{\alpha}{2}}\hat{N}^{\frac{\alpha}{2}-1}\frac{\tilde{r}_{\max}}{\tilde{r}_{\min}}$, which means }its worst-case performance for minimizing the SC platform's crowdsourcing cost can guarantee $
		\hat{l}{}_{m}(\mathsf{M}_{\mathrm{MED}}(\boldsymbol{L}))\leq2^{\frac{\alpha}{2}}\hat{N}^{\frac{\alpha}{2}-1}\frac{\tilde{r}_{\max}}{\tilde{r}_{\min}}\hat{l}{}_{m}(\mathsf{M}_{\mathrm{OPT}}(\boldsymbol{L}))$.
\end{prop}
\begin{IEEEproof}
We expand the SC platform's utility function in~(\ref{eq:SC_utility_data_crowdsourcing})
as follows:
\begin{align}
 & \hat{l}{}_{m}((x_{\mathrm{M}},y_{\mathrm{M}}))\nonumber \\
 & =\frac{\tilde{P}_{c}\kappa}{\sum_{j\in\hat{\mathcal{N}}}\tilde{r}_{j}}\sum_{i\in\hat{\mathcal{N}}}\left(\tilde{r}_{i}^{\frac{2}{\alpha}}(x_{i}-x_{\mathrm{M}})^{2}+(y_{i}-y_{\mathrm{M}})^{2}+h^{2}\right)^{\frac{\alpha}{2}}.\label{eq:SC-utility-rewitten-DCP}
\end{align}
Let $x_{\mathrm{med}},\overline{x}$ and $y_{\mathrm{med}},\overline{y}$ respectively denote the median and mean of $\mathbf{x}=(x_{1},\ldots,x_{\hat{N}})$ and $\mathbf{y}=(y_{1},\ldots,y_{\hat{N}})$. Also, we use $(x_{\mathrm{opt}},y_{\mathrm{opt}})$ to denote the optimal solution to maximizing the utility function in (\ref{eq:SC-utility-rewitten-DCP}), i.e., $\mathsf{M}_{\mathrm{OPT}}(\boldsymbol{L})=(x_{\mathrm{opt}},y_{\mathrm{opt}})$. We also note that the optimal solution to minimizing the $\sum_{i\in\hat{\mathcal{N}}}\tilde{r}_{i}^{\frac{2}{\alpha}}\left((x_{i}-x_{\mathrm{M}})^{2}+(y_{i}-y_{\mathrm{M}})^{2}+h^{2}\right)$ is $(x^{*},y^{*})$ where $x^{*}=\frac{\sum_{i\in\hat{\mathcal{N}}}\tilde{r}_{i}^{\frac{2}{\alpha}}x_{i}}{\sum_{i\in\hat{\mathcal{N}}}\tilde{r}_{i}^{\frac{2}{\alpha}}}$ and $y^{*}=\frac{\sum_{i\in\hat{\mathcal{N}}}\tilde{r}_{i}^{\frac{2}{\alpha}}y_{i}}{\sum_{i\in\hat{\mathcal{N}}}\tilde{r}_{i}^{\frac{2}{\alpha}}}$. As $\tilde{r}_{\min}\leq\tilde{r}_{i}$, we have
\begin{align}
 & \tilde{r}_{\min}^{\frac{2}{\alpha}}\sum_{i\in\hat{\mathcal{N}}}\left((x_{i}-\overline{x})^{2}+(y_{i}-\overline{y})^{2}+h^{2}\right)\nonumber \\
 & \leq\sum_{i\in\hat{\mathcal{N}}}\tilde{r}_{i}^{\frac{2}{\alpha}}\left((x_{i}-x^{*})^{2}+(y_{i}-y^{*})^{2}+h^{2}\right).
\end{align}
According to~\cite[Theorem 4.3]{Feldman2013}, we have $\sum_{i\in\hat{\mathcal{N}}}(x_{i}-x_{\mathrm{med}})^{2}\leq2\sum_{i\in\hat{\mathcal{N}}}(x_{i}-\overline{x})^{2}$ and $\sum_{i\in\hat{\mathcal{N}}}(y_{i}-y_{\mathrm{med}})^{2}\leq2\sum_{i\in\hat{\mathcal{N}}}(y_{i}-\overline{y})^{2}$. Then, we can verify that 
\begin{align}
 & \tilde{r}_{\min}^{\frac{2}{\alpha}}\sum_{i\in\hat{\mathcal{N}}}\left((x_{i}-x_{\mathrm{med}})^{2}+(y_{i}-y_{\mathrm{med}})^{2}+h^{2}\right)\nonumber \\
 & \leq2\tilde{r}_{\min}^{\frac{2}{\alpha}}\sum_{i\in\hat{\mathcal{N}}}\left((x_{i}-\overline{x})^{2}+(y_{i}-\overline{y})^{2}+h^{2}\right),\\
 & \tilde{r}_{\min}\left(\sum_{i\in\hat{\mathcal{N}}}\left((x_{i}-x_{\mathrm{med}})^{2}+(y_{i}-y_{\mathrm{med}})^{2}+h^{2}\right)\right)^{\frac{\alpha}{2}}\nonumber \\
 & \leq2^{\frac{\alpha}{2}}\tilde{r}_{\min}\left(\sum_{i\in\hat{\mathcal{N}}}\left((x_{i}-\overline{x})^{2}+(y_{i}-\overline{y})^{2}+h^{2}\right)\right)^{\frac{\alpha}{2}}\nonumber \\
 & \leq2^{\frac{\alpha}{2}}\tilde{r}_{\min}\left(\sum_{i\in\hat{\mathcal{N}}}\left((x_{i}-x^{*})^{2}+(y_{i}-y^{*})^{2}+h^{2}\right)\right)^{\frac{\alpha}{2}}\nonumber \\
 & \leq2^{\frac{\alpha}{2}}\left(\sum_{i\in\hat{\mathcal{N}}}\tilde{r}_{i}^{\frac{2}{\alpha}}\left((x_{i}-x^{*})^{2}+(y_{i}-y^{*})^{2}+h^{2}\right)\right)^{\frac{\alpha}{2}}.\label{eq:inquality-med-mean}
\end{align}
Since $\alpha\geq2$, we can prove that
\begin{align}
\tilde{r}_{\min}\sum_{i\in\hat{\mathcal{N}}}\left((x_{i}-x_{\mathrm{med}})^{2}+(y_{i}-y_{\mathrm{med}})^{2}+h^{2}\right)^{\frac{\alpha}{2}}\nonumber \\
\leq\tilde{r}_{\min}\left(\sum_{i\in\mathcal{\hat{N}}}\left((x_{i}-x_{\mathrm{med}})^{2}+(y_{i}-y_{\mathrm{med}})^{2}+h^{2}\right)\right)^{\frac{\alpha}{2}}.
\end{align}
Hence, based on Theorem $1$ in~\cite{Jameson2014} and the fact that $\tilde{r}_{i}\leq\tilde{r}_{\max}$ and $\frac{\alpha}{2}\geq1$, we can obtain
\begin{align}
 & 2^{\frac{\alpha}{2}}\left(\sum_{i\in\hat{\mathcal{N}}}\tilde{r}_{i}^{\frac{2}{\alpha}}\left((x_{i}-x^{*})^{2}+(y_{i}-y^{*})^{2}+h^{2}\right)\right)^{\frac{\alpha}{2}}\nonumber \\
 & \leq2^{\frac{\alpha}{2}}\left(\sum_{i\in\hat{\mathcal{N}}}\tilde{r}_{i}^{\frac{2}{\alpha}}\left((x_{i}-x_{\mathrm{opt}})^{2}+(y_{i}-y_{\mathrm{opt}})^{2}+h^{2}\right)\right)^{\frac{\alpha}{2}}\nonumber \\
 & \leq2^{\frac{\alpha}{2}}\tilde{r}_{\max}\left(\sum_{i\in\hat{\mathcal{N}}}\left((x_{i}-x_{\mathrm{opt}})^{2}+(y_{i}-y_{\mathrm{opt}})^{2}+h^{2}\right)\right)^{\frac{\alpha}{2}}\nonumber \\
 & \leq2^{\frac{\alpha}{2}}\hat{N}^{\frac{\alpha}{2}-1}\tilde{r}_{\max}\sum_{i\in\hat{\mathcal{N}}}\left((x_{i}-x_{\mathrm{opt}})^{2}+(y_{i}-y_{\mathrm{opt}})^{2}+h^{2}\right)^{\frac{\alpha}{2}}.
\end{align}
Combining the above inequalities, we have
\begin{align*}
 & \frac{\tilde{P}_{c}\kappa}{\sum_{j\in\hat{\mathcal{N}}}\tilde{r}_{j}}\sum_{i\in\hat{\mathcal{N}}}\left((x_{i}-x_{\mathrm{med}})^{2}+(y_{i}-y_{\mathrm{med}})^{2}+h^{2}\right)^{\frac{\alpha}{2}}
\end{align*}
\begin{align*}
 & \leq2^{\frac{\alpha}{2}}\hat{N}^{\frac{\alpha}{2}-1}\frac{\tilde{r}_{\max}}{\tilde{r}_{\min}}\frac{\tilde{P}_{c}\kappa}{\sum_{j\in\hat{\mathcal{N}}}\tilde{r}_{j}}\\
 & \quad\quad\sum_{i\in\hat{\mathcal{N}}}\left((x_{i}-x_{\mathrm{opt}})^{2}+(y_{i}-y_{\mathrm{opt}})^{2}+h^{2}\right)^{\frac{\alpha}{2}}.
\end{align*}
Finally, we can conclude that $\hat{l}{}_{m}(\mathsf{M}_{\mathrm{MED}}(\boldsymbol{L}))\leq2^{\frac{\alpha}{2}}\hat{N}^{\frac{\alpha}{2}-1}\frac{\tilde{r}_{\max}}{\tilde{r}_{\min}}\hat{l}{}_{m}(\mathsf{M}_{\mathrm{OPT}}(\boldsymbol{L}))$.
\end{IEEEproof}

However, we find that the MED mechanism can be arbitrarily inefficient, especially when the wireless channel path-loss and the number of workers are large. Thanks to the workers' historical location data kept by the SC platform, it is possible to design mechanisms that achieve higher utility \emph{in expectation}. Each worker's location $(x_{i},y_{i})$ follows a distribution whose joint continuous probability density function (PDF) is $\mathcal{P}_{i}$ on its working area $\boldsymbol{A}_{i}$, i.e., $(x_{i},y_{i})\sim\mathcal{P}_{i}$ for $i=1,\ldots,\hat{N}$. With a slight abuse of notation, let the probability density function of $\mathcal{P}_{i}$ at a pair of real numbers $(x_{i},y_{i})$ be $\mathcal{P}_{i}(x_{i},y_{i})$. Under the Bayesian setting, we propose an enhanced median-single-constant (MSC) mechanism (shown in Algorithm~\ref{alg:mean-median-mechanism}) where we add a single constant point $(x_{\mathrm{c}},y_{\mathrm{c}})$ in the original set of input locations and then run the median mechanism on the new set. According to Theorems~\ref{thm:(Moulin's-one-dimensional-genera)} and~\ref{thm:Multi-dimensional-generalized-median}, the MSC mechanism is equivalent to respectively setting one constant on each dimension at a fixed value while setting the half of the other constants at the positive infinity and the remaining half at the negative infinity. Hence, its design rationale follows the multi-dimensional generalized median mechanism and is strategyproof. We obtain vectors $\mathbf{x}=(x_{1},\ldots,x_{\hat{N}})$ and $\mathbf{y}=(y_{1},\ldots,y_{\hat{N}})$ from $\mathbf{L}=((x_{1},y_{1}),\ldots,(x_{i},y_{i}),\ldots,(x_{\hat{N}},y_{\hat{N}}))$. Let $(x_{\mathrm{med}},y_{\mathrm{med}})=\mathsf{M}_{\mathrm{MED}}(\mathbf{L})$ and $(x_{\mathrm{msc}},y_{\mathrm{msc}})=\mathsf{M}_{\mathrm{MSC}}(\mathbf{L})$ respectively be the outcome from the MED mechanism and the MSC mechanism. Next, we analyze their expected performance.
\begin{algorithm}[tbh]
\scriptsize
\begin{algorithmic}[1] 
\Require{Workers' reported locations $\mathbf{L}=(L_1,\ldots, L_i, \ldots,L_{\hat{N}})$ where $L_i=(x_i,y_i), i\in{\mathcal{\hat{N}}}$ and the worker's location  distribution $\mathcal{P}_{i}(x_{i},y_{i}), i\in{\mathcal{\hat{N}}}$.}
\Ensure{Mobile BS's service location $L_M=(x_{\mathrm{M}},y_{\mathrm{M}})$.}
\Begin
	\State{Calculate $x_\mathrm{c}$ and $y_\mathrm{c}$ based on $\mathcal{P}_{i}(x_{i},y_{i}), i\in{\hat{\mathcal{N}}}$.}
	\State{Add the constant point $(x_\mathrm{c},y_\mathrm{c})$ to $\mathbf{L}$, i.e., $\mathbf{L}_{\mathrm{c}} \gets \mathbf{L} \cup (x_\mathrm{c},y_\mathrm{c})$.}
	\State{Run the median mechanism on the new $\mathbf{L}_{\mathrm{c}}$ (${\hat{N}}+1$ location points) and output the $x_{\mathrm{M}}$ and $y_{\mathrm{M}}$.}
\End
\end{algorithmic} 

\caption{MSC mechanism\label{alg:mean-median-mechanism}}
\end{algorithm}
With $\mathbb{E}[\cdot]$ denoting the expectation, for the SC platform's
crowdsourcing cost $\hat{l}{}_{m}$ in (\ref{eq:SC_cost_function}),
we compute
\begin{align}
 & \mathbb{E}_{(x_{i},y_{i})\sim\mathcal{P}^{i},\,i\in\hat{\mathcal{N}}}[\hat{l}{}_{m}(\mathsf{M}_{\mathrm{MED}}(\mathbf{L}))]\nonumber \\
 & =\iint{}_{(x_{1},y_{1})\in\boldsymbol{A}_{1}}\cdots\iint{}_{(x_{\hat{N}},y_{\hat{N}})\in\boldsymbol{A}_{\hat{N}}}\hat{l}{}_{m}(\mathsf{M}_{\mathrm{MED}}(\mathbf{L}))\nonumber \\
 & \;\;\;\;\mathcal{P}_{1}(x_{1},y_{1})\cdots\mathcal{P}_{N}(x_{\hat{N}},y_{\hat{N}})\,\mathrm{d}x_{1}\cdots\mathrm{d}x_{\hat{N}}\mathrm{d}y_{1}\cdots\mathrm{d}y_{\hat{N}}.\label{eq:SC_expected_cost_1}
\end{align}
For ease of the analysis, we assume that all workers are independently and identically distributed following the same continuous PDF $\mathcal{P}$ on the domain $\boldsymbol{A}$ in the rest of the subsection. In order to simplify the operation with symmetry, we first define and investigate $\hat{l}{}_{m}(\mathsf{M}_{\mathrm{MED}}(\mathbf{L}))$ by setting each worker's transmission rate $\tilde{r}_{i}=1$. As the PDF is continuous, we consider only the case where $x_{1},\ldots,x_{\hat{N}},y_{1},\ldots,y_{\hat{N}}$ are all different. When $x_{1},\ldots,x_{\hat{N}},y_{1},\ldots,y_{\hat{N}}$ are all different, to sort $x_{1},\ldots,x_{\hat{N}}$ and $y_{1},\ldots,y_{\hat{N}}$ in ascending order, we have $(\hat{N}!)^{2}$ possibilities. Hence, it follows that 
\begin{align}
 & \mathbb{E}_{(x_{i},y_{i})\sim\mathcal{P},\,i\in\hat{\mathcal{N}}}[\hat{l}{}_{m}(\mathsf{M}_{\mathrm{MED}}(\mathbf{L}))]\\&=\iint_{(x_{1},y_{1})\in\boldsymbol{A}}\cdots\iint_{(x_{\hat{N}},y_{\hat{N}})\in\boldsymbol{A}}\hat{l}{}_{m}(\mathsf{M}_{\mathrm{MED}}(\mathbf{L}))\nonumber \\
 & \mathcal{P}(x_{1},y_{1})\cdots\mathcal{P}(x_{\hat{N}},y_{\hat{N}})\,\mathrm{d}x_{1}\cdots\mathrm{d}x_{\hat{N}}\mathrm{d}y_{1}\cdots\mathrm{d}y_{\hat{N}}\nonumber \\
 & =(\hat{N}!)^{2}\iint_{\begin{subarray}{l}
(x_{1},y_{1}),\ldots,(x_{\hat{N}},y_{\hat{N}})\in\boldsymbol{A}\\
x_{1}<\cdots<x_{\hat{N}}\\
y_{1}<\cdots<y_{\hat{N}}
\end{subarray}}\hat{l}{}_{m}(\mathsf{M}_{\mathrm{MED}}(\mathbf{L}))\nonumber \\
 & \,\,\,\,\,\,\,\,\,\,\,\mathcal{P}(x_{1},y_{1})\cdots\mathcal{P}(x_{\hat{N}},y_{\hat{N}})\,\mathrm{d}x_{1}\cdots\mathrm{d}x_{\hat{N}}\mathrm{d}y_{1}\cdots\mathrm{d}y_{\hat{N}}.\label{eq:SC_expected_cost_2}
\end{align}
Given $x_{1}<x_{2}<\cdots<x_{\hat{N}}$ and $y_{1}<y_{2}<\cdots<y_{\hat{N}}$, we can have $\mathsf{M}_{\mathrm{MED}}(\mathbf{L})=(\frac{x_{\frac{\hat{N}}{2}}+x_{\frac{\hat{N}}{2}+1}}{2},\frac{y_{\frac{\hat{N}}{2}}+y_{\frac{\hat{N}}{2}+1}}{2})$ for even $\hat{N}$ and $\mathsf{M}_{\mathrm{MED}}(\mathbf{L})=(x_{\frac{\hat{N}+1}{2}},y_{\frac{\hat{N}+1}{2}})$ for odd $\hat{N}$ according to the MED mechanism (Algorithm~\ref{alg:MED-mechanism}). After substituting the expression of $\mathsf{M}_{\mathrm{MED}}(\mathbf{L})$ into equation (\ref{eq:SC_expected_cost_2}), we can combine equations (\ref{eq:SC_expected_cost_1}) and (\ref{eq:SC_expected_cost_2}) to obtain 
\begin{align}
 & \mathbb{E}_{(x_{i},y_{i})\sim\mathcal{P},\,i\in\hat{\mathcal{N}}}[\hat{l}{}_{m}(\mathsf{M}_{\mathrm{MED}}(\mathbf{L}))]\nonumber \\
 & =\begin{cases}
(\hat{N}!)^{2}\iint_{\begin{subarray}{l}
(x_{1},y_{1}),\ldots,(x_{\hat{N}},y_{\hat{N}})\in\boldsymbol{A}\\
x_{1}<\cdots<x_{\hat{N}}\\
y_{1}<\cdots<y_{\hat{N}}
\end{subarray}}\hat{l}{}_{m}((\frac{x_{\frac{\hat{N}}{2}}+x_{\frac{\hat{N}}{2}+1}}{2},\frac{y_{\frac{\hat{N}}{2}}+y_{\frac{\hat{N}}{2}+1}}{2}))\\
\,\,\,\,\,\,\,\,\,\,\,\,\,\mathcal{P}(x_{1},y_{1})\cdots\mathcal{P}(x_{\hat{N}},y_{\hat{N}})\,\mathrm{d}x_{1}\cdots\mathrm{d}x_{\hat{N}}\mathrm{d}y_{1}\cdots\mathrm{d}y_{\hat{N}},\\
\,\,\,\,\,\,\,\,\,\,\,\,\,\,\,\,\,\,\,\,\,\,\,\,\,\,\,\,\,\,\,\,\,\,\,\,\,\,\,\,\,\,\,\,\,\,\,\,\,\,\,\,\,\,\,\,\,\,\,\,\,\,\,\,\,\,\,\,\,\,\,\,\,\,\,\,\,\,\,\,\,\,\,\,\,\,\,\,\,\,\,\,\,\,\,\,\,\,\,\,\,\,\,\,\,for\,even\,\hat{N},\\
(\hat{N}!)^{2}\iint_{\begin{subarray}{l}
(x_{1},y_{1}),\ldots,(x_{\hat{N}},y_{\hat{N}})\in\boldsymbol{A}\\
x_{1}<\cdots<x_{\hat{N}}\\
y_{1}<\cdots<y_{\hat{N}}
\end{subarray}}\hat{l}{}_{m}((x_{\frac{\hat{N}+1}{2}},y_{\frac{\hat{N}+1}{2}}))\\
\,\,\,\,\,\,\,\,\,\,\,\,\mathcal{P}(x_{1},y_{1})\cdots\mathcal{P}(x_{\hat{N}},y_{\hat{N}})\,\mathrm{d}x_{1}\cdots\mathrm{d}x_{\hat{N}}\mathrm{d}y_{1}\cdots\mathrm{d}y_{\hat{N}},\\
\,\,\,\,\,\,\,\,\,\,\,\,\,\,\,\,\,\,\,\,\,\,\,\,\,\,\,\,\,\,\,\,\,\,\,\,\,\,\,\,\,\,\,\,\,\,\,\,\,\,\,\,\,\,\,\,\,\,\,\,\,\,\,\,\,\,\,\,\,\,\,\,\,\,\,\,\,\,\,\,\,\,\,\,\,\,\,\,\,\,\,\,\,\,\,\,\,\,\,\,\,\,\,\,for\,odd\,\hat{N}.
\end{cases}\label{eq:SC_Cost_expectation_3}
\end{align}
Then, considering the symmetry of each worker, we can use (\ref{eq:SC_Cost_expectation_3}) to obtain the simplified expression of (\ref{eq:SC_expected_cost_1}) as follows:
\begin{align}
 & \mathbb{E}_{(x_{i},y_{i})\sim\mathcal{P},\,i\in\hat{\mathcal{N}}}[\hat{l}{}_{m}(\mathsf{M}_{\mathrm{MED}}(\mathbf{L}))]\nonumber \\
 & =\frac{1}{\hat{N}}\sum_{j\in\hat{\mathcal{N}}}\tilde{r}_{j}\mathbb{E}_{(x_{i},y_{i})\sim\mathcal{P},\,i\in\hat{\mathcal{N}}}[\hat{l}{}_{m}(\mathsf{M}_{\mathrm{MED}}(\mathbf{L}))].\label{eq:SC-uitility-MED-simplified-1}
\end{align} 

For the MSC mechanism, we study its performance in a similar way as above and address the problem of how to calculate the constant point $(x_{\mathrm{c}},y_{\mathrm{c}})$ by leveraging the known distribution $\mathcal{P}$. Due to the symmetry and the limited space, the following analysis just shows cases where $x_{1}<x_{2}<\cdots<x_{\hat{N}}$, $y_{1}<y_{2}<\cdots<y_{\hat{N}}$, $\hat{N}$ is odd and the $x_{\mathrm{c}}$ is smaller than $x_{\frac{\hat{N}-1}{2}}$, i.e., $x_{\mathrm{c}}\leq x_{\frac{\hat{N}-1}{2}}$. It can be extended for other cases where $N$ is even and $x_{\mathrm{c}}$ is more general. 

1) Case 1: When $x_{\mathrm{c}}<x_{\frac{\hat{N}-1}{2}}$ and $y_{\mathrm{c}}<y_{\frac{\hat{N}-1}{2}}$,
then $\mathsf{M}_{\mathrm{MSC}}(\mathbf{L})=(\frac{x_{\frac{\hat{N}-1}{2}}+x_{\frac{\hat{N}+1}{2}}}{2},\frac{y_{\frac{\hat{N}-1}{2}}+y_{\frac{\hat{N}+1}{2}}}{2})$
and
\begin{align}
 & \mathbb{E}_{(x_{i},y_{i})\sim\mathcal{P},\,i\in\hat{\mathcal{N}}}[\hat{l}{}_{m}(\mathsf{M}_{\mathrm{MSC}}(\mathbf{L}))]\nonumber \\
 & =\iint_{\begin{subarray}{l}
(x_{1},y_{1}),\ldots,(x_{\hat{N}},y_{\hat{N}})\in\boldsymbol{A}\\
x_{\frac{\hat{N}-3}{2}}<x_{\mathrm{c}}<x_{\frac{\hat{N}-1}{2}}\\
y_{\frac{\hat{N}-3}{2}}<y_{\mathrm{c}}<y_{\frac{\hat{N}-1}{2}}
\end{subarray}}\hat{l}{}_{m}((\frac{x_{\frac{\hat{N}-1}{2}}+x_{\frac{\hat{N}+1}{2}}}{2},\frac{y_{\frac{\hat{N}-1}{2}}+y_{\frac{\hat{N}+1}{2}}}{2}))\nonumber \\
 & \mathcal{P}(x_{1},y_{1})\cdots\mathcal{P}(x_{\hat{N}},y_{\hat{N}})\,\mathrm{d}x_{1}\cdots\mathrm{d}x_{\hat{N}}\mathrm{d}y_{1}\cdots\mathrm{d}y_{\hat{N}}\underset{((\frac{\hat{N}-1}{2})^{2}-2)\,terms}{\underbrace{+\cdots+}}\nonumber \\
 & \iint_{\begin{subarray}{l}
(x_{1},y_{1}),\ldots,(x_{\hat{N}},y_{\hat{N}})\in\boldsymbol{A}\\
x_{c}<x_{1}\\
y_{c}<y_{1}
\end{subarray}}\hat{l}{}_{m}((\frac{x_{\frac{\hat{N}-1}{2}}+x_{\frac{\hat{N}+1}{2}}}{2},\frac{y_{\frac{\hat{N}-1}{2}}+y_{\frac{\hat{N}+1}{2}}}{2}))\nonumber \\
 & \mathcal{P}(x_{1},y_{1})\cdots\mathcal{P}(x_{\hat{N}},y_{\hat{N}})\,\mathrm{d}x_{1}\cdots\mathrm{d}x_{\hat{N}}\mathrm{d}y_{1}\cdots\mathrm{d}y_{\hat{N}}.\label{eq:Case1-msc-mechanism}
\end{align}

2) Case 2: When $x_{\mathrm{c}}<x_{\frac{\hat{N}-1}{2}}$ and $y_{\frac{\hat{N}-1}{2}}<y_{\mathrm{c}}<y_{\frac{\hat{N}+3}{2}}$,
then $\mathsf{M}_{\mathrm{MSC}}(\mathbf{L})=(\frac{x_{\frac{\hat{N}-1}{2}}+x_{\frac{\hat{N}+1}{2}}}{2},\frac{y_{\mathrm{c}}+y_{\frac{\hat{N}+1}{2}}}{2})$
and

\begin{align}
 & \mathbb{E}_{(x_{i},y_{i})\sim\mathcal{P},\,i\in\hat{\mathcal{N}}}[\hat{l}{}_{m}(\mathsf{M}_{\mathrm{MSC}}(\mathbf{L}))]\nonumber \\
 & =\iint_{\begin{subarray}{l}
(x_{1},y_{1}),\ldots,(x_{N},y_{N})\in\boldsymbol{A}\\
x_{\frac{\hat{N}-3}{2}}<x_{\mathrm{c}}<x_{\frac{\hat{N}-1}{2}}\\
y_{\frac{\hat{N}-1}{2}}<y_{\mathrm{c}}<y_{\frac{\hat{N}+3}{2}}
\end{subarray}}\hat{l}{}_{m}((\frac{x_{\frac{\hat{N}-1}{2}}+x_{\frac{\hat{N}+1}{2}}}{2},\frac{y_{\mathrm{c}}+y_{\frac{\hat{N}+1}{2}}}{2}))\nonumber \\
 & \mathcal{P}(x_{1},y_{1})\cdots\mathcal{P}(x_{\hat{N}},y_{\hat{N}})\,\mathrm{d}x_{1}\cdots\mathrm{d}x_{\hat{N}}\mathrm{d}y_{1}\cdots\mathrm{d}y_{\hat{N}}\underset{(\hat{N}-3)\,terms}{\underbrace{+\cdots+}}\nonumber \\
 & \iint_{\begin{subarray}{l}
(x_{1},y_{1}),\ldots,(x_{\hat{N}},y_{\hat{N}})\in\boldsymbol{A}\\
x_{c}<x_{1}\\
y_{\frac{\hat{N}+1}{2}}<y_{c}<y_{\frac{\hat{N}+3}{2}}
\end{subarray}}\hat{l}{}_{m}((\frac{x_{\frac{\hat{N}-1}{2}}+x_{\frac{\hat{N}+1}{2}}}{2},\frac{y_{\mathrm{c}}+y_{\frac{\hat{N}+1}{2}}}{2}))\nonumber \\
 & \mathcal{P}(x_{1},y_{1})\cdots\mathcal{P}(x_{\hat{N}},y_{\hat{N}})\,\mathrm{d}x_{1}\cdots\mathrm{d}x_{\hat{N}}\mathrm{d}y_{1}\cdots\mathrm{d}y_{\hat{N}}.\label{eq:Case2-msc-mechanism}
\end{align}

3) Case 3: When $x_{\mathrm{c}}<x_{\frac{\hat{N}-1}{2}}$ and $y_{\frac{\hat{N}+3}{2}}<y_{\mathrm{c}}$,
then $\mathsf{M}_{\mathrm{MSC}}(\mathbf{L})=(\frac{x_{\frac{\hat{N}-1}{2}}+x_{\frac{\hat{N}+1}{2}}}{2},\frac{y_{\frac{\hat{N}+1}{2}}+y_{\frac{\hat{N}+3}{2}}}{2})$
and

\begin{align}
 & \mathbb{E}_{(x_{i},y_{i})\sim\mathcal{P},\,i\in\hat{\mathcal{N}}}[\hat{l}{}_{m}(\mathsf{M}_{\mathrm{MSC}}(\mathbf{L}))]\nonumber \\
 & =\iint_{\begin{subarray}{l}
(x_{1},y_{1}),\ldots,(x_{\hat{N}},y_{\hat{N}})\in\boldsymbol{A}\\
x_{\frac{\hat{N}-3}{2}}<x_{\mathrm{c}}<x_{\frac{\hat{N}-1}{2}}\\
y_{\frac{\hat{N}+3}{2}}<y_{\mathrm{c}}<y_{\frac{\hat{N}+5}{2}}
\end{subarray}}\hat{l}{}_{m}((\frac{x_{\frac{\hat{N}-1}{2}}+x_{\frac{\hat{N}+1}{2}}}{2},\frac{y_{\frac{\hat{N}+1}{2}}+y_{\frac{\hat{N}+3}{2}}}{2}))\nonumber \\
 & \mathcal{P}(x_{1},y_{1})\cdots\mathcal{P}(x_{\hat{N}},y_{\hat{N}})\,\mathrm{d}x_{1}\cdots\mathrm{d}x_{\hat{N}}\mathrm{d}y_{1}\cdots\mathrm{d}y_{\hat{N}}\underset{((\frac{\hat{N}-1}{2})^{2}-2)\,terms}{\underbrace{+\cdots+}}\nonumber \\
 & \iint_{\begin{subarray}{l}
(x_{1},y_{1}),\ldots,(x_{\hat{N}},y_{\hat{N}})\in\boldsymbol{A}\\
x_{\mathrm{c}}<x_{1}\\
y_{\hat{N}}<y_{\mathrm{c}}
\end{subarray}}\hat{l}{}_{m}((\frac{x_{\frac{\hat{N}-1}{2}}+x_{\frac{\hat{N}+1}{2}}}{2},\frac{y_{\frac{\hat{N}+1}{2}}+y_{\frac{\hat{N}+3}{2}}}{2}))\nonumber \\
 & \mathcal{P}(x_{1},y_{1})\cdots\mathcal{P}(x_{\hat{N}},y_{\hat{N}})\,\mathrm{d}x_{1}\cdots\mathrm{d}x_{\hat{N}}\mathrm{d}y_{1}\cdots\mathrm{d}y_{\hat{N}}.\label{eq:Case3-msc-mechanism}
\end{align}

There are totally $((\hat{N}+1)!)^{2}$ terms similar to (\ref{eq:Case3-msc-mechanism}) to compute the expected utility achieved by the MSC mechanism, which is challenging especially when $\hat{N}$ is large. Next, we would like to present a special case to show the possibility and feasibility to maximize the expected utility through optimizing $(x_{\mathrm{c}},y_{\mathrm{c}})$. In the special case, we assume that each worker's location follows the bivariate uniform distribution, i.e., $\mathcal{P}^{\mathrm{u}}=\begin{cases} 1, & (x,y)\in\boldsymbol{A}=[0,1]^{2},\\ 0, & otherwise, \end{cases}$ and the path-loss $\alpha$ is $2$. Then, by substituting these parameters into (\ref{eq:SC_expected_cost_1})-(\ref{eq:SC_Cost_expectation_3}) and using mathematical induction, we first obtain the expected utility generated by the MED mechanism as 
\begin{align}
 & \mathbb{E}_{(x_{i},y_{i})\sim\mathcal{P}^{\mathrm{u}},\,i\in\hat{\mathcal{N}}}[\hat{l}{}_{m}(\mathsf{M}_{\mathrm{MED}}(\mathbf{L}))]\nonumber \\
 & =\begin{cases}
\tilde{P}_{c}\kappa\left(\frac{(N-1)(N+4)}{6(N+1)(N+2)}+h^{2}\right), & for\,even\,\hat{N},\\
\tilde{P}_{c}\kappa\left(\frac{(N-1)(N+3)}{6N(N+2)}+h^{2}\right), & for\,odd\,\hat{N}.
\end{cases}\label{eq:expected-utility-derived-median-mechanism}
\end{align}
For the MSC mechanism, we analyze a situation where there are three employed workers, i.e., $\hat{N}=3$. The same way of the analysis can be applied to any number of employed workers. Based on the analysis above, we can calculate the expected utility achieved by the MSC mechanism as follows:
\begin{align}
 & \mathbb{E}_{(x_{i},y_{i})\sim\mathcal{P},\,i\in\{1,2,3\}}[\hat{u}{}_{m}(\mathsf{M}_{\mathrm{MSC}}(\mathbf{L}))]\nonumber \\
 & =\tilde{P}_{c}\kappa\left(\underset{\text{\ding{192}}}{\underbrace{-\frac{x_{\mathrm{c}}^{4}+y_{\mathrm{c}}^{4}}{4}+\frac{x_{\mathrm{c}}^{3}+y_{\mathrm{c}}^{3}}{2}-\frac{x_{\mathrm{c}}^{2}+y_{\mathrm{c}}^{2}}{4}+\frac{3}{20}}}+h^{2}\right).\label{eq:SC-expected-utility-N-MSC-3}
\end{align}
Then, the expected utility achieved by the MED mechanism is 
\begin{equation}
\mathbb{E}_{(x_{i},y_{i})\sim\mathcal{P},\,i\in\{1,2,3\}}[\hat{u}{}_{m}(\mathsf{M}_{\mathrm{MED}}(\mathbf{L}))]=\tilde{P}_{c}\kappa\left(\frac{2}{15}+h^{2}\right).
\end{equation}
The minimum value of $\text{\ding{192}}$ in equation (\ref{eq:SC-expected-utility-N-MSC-3}) is $\frac{19}{160}$ achieved at $x_{\mathrm{c}}=y_{\mathrm{c}}=0.5$, which is smaller than $\frac{2}{15}$. Hence, we can find a constant point $(x_{\mathrm{c}},y_{\mathrm{c}})$ that enables the MSC mechanism to achieve lower expected crowdsourcing cost than that of the benchmark MED mechanism. This also indicates the possibility of improving and extending the MSC mechanism for more general scenarios.

\subsection{Deep learning based mobile BS deployment mechanism \label{sec:Deep-learning-based} }

Clearly, above conventional mechanisms, including the MED and MSC mechanism, have several non-negligible limitations: 
\begin{itemize}
\item It is intractable to manually optimize the MSC mechanism in realistic
environments where path-loss exponent $\alpha$ is not necessarily
$2$ and the number of employed workers $\hat{N}$ may be much larger
than $3$. 
\item Each worker's working location distribution can be different and correlated.
Despite the location distribution can be inferred from historical
data, its accurate type is not always known or even there is no corresponding
closed-form expression for us to proceed with the theoretical analysis.
\item In the MSC mechanism, only a single constant point is optimized while
the generalized median mechanism implies that more constant points
can be used to improve the expected performance. 
\end{itemize}
To overcome the above limitations, we develop a deep learning based mechanism named the MDL mechanism. The MDL mechanism provides an efficient model-free method to simultaneously exploit the data and optimize the complicated objective utility function while satisfying the incentive constraints. In the construction of the deep neural network, we use an equivalent definition (Theorem~\ref{thm:(Molin's-max-min-rule)}) of the one-dimensional generalized median mechanism (Theorem~\ref{thm:(Moulin's-one-dimensional-genera)}). 
\begin{thm}
\label{thm:(Molin's-max-min-rule)}(\cite{Moulin1980,Border1983,Nisan2007})
A mechanism $\mathsf{M}$ is a strategyproof and anonymous generalized
median mechanism on one dimensional space if there exist $2^{\hat{N}}$
points $\left\{ \zeta_{\mathcal{T}}\right\} _{\mathcal{T}\subseteq\hat{\mathcal{N}}}$
in $[\zeta_{\varnothing},\zeta_{\mathcal{\hat{N}}}]$, such that 1)
$\mathcal{T}\subseteq\mathcal{T}'\subseteq\hat{\mathcal{N}}$ implies
$\zeta_{\mathcal{T}}\le\zeta_{\mathcal{T}'}$ and 2) for all $\mathbf{x}\in\mathbb{R}^{\hat{N}}$,
$\mathsf{M}(\mathbf{x})=\max_{\mathcal{T}\subseteq\hat{\mathcal{N}}}\min\left\{ \zeta_{\mathcal{T}},\,x_{i}:i\in\mathcal{T}\right\} $.
\end{thm}
\begin{figure}[tbh]
\begin{centering}
\includegraphics[width=0.8\columnwidth]{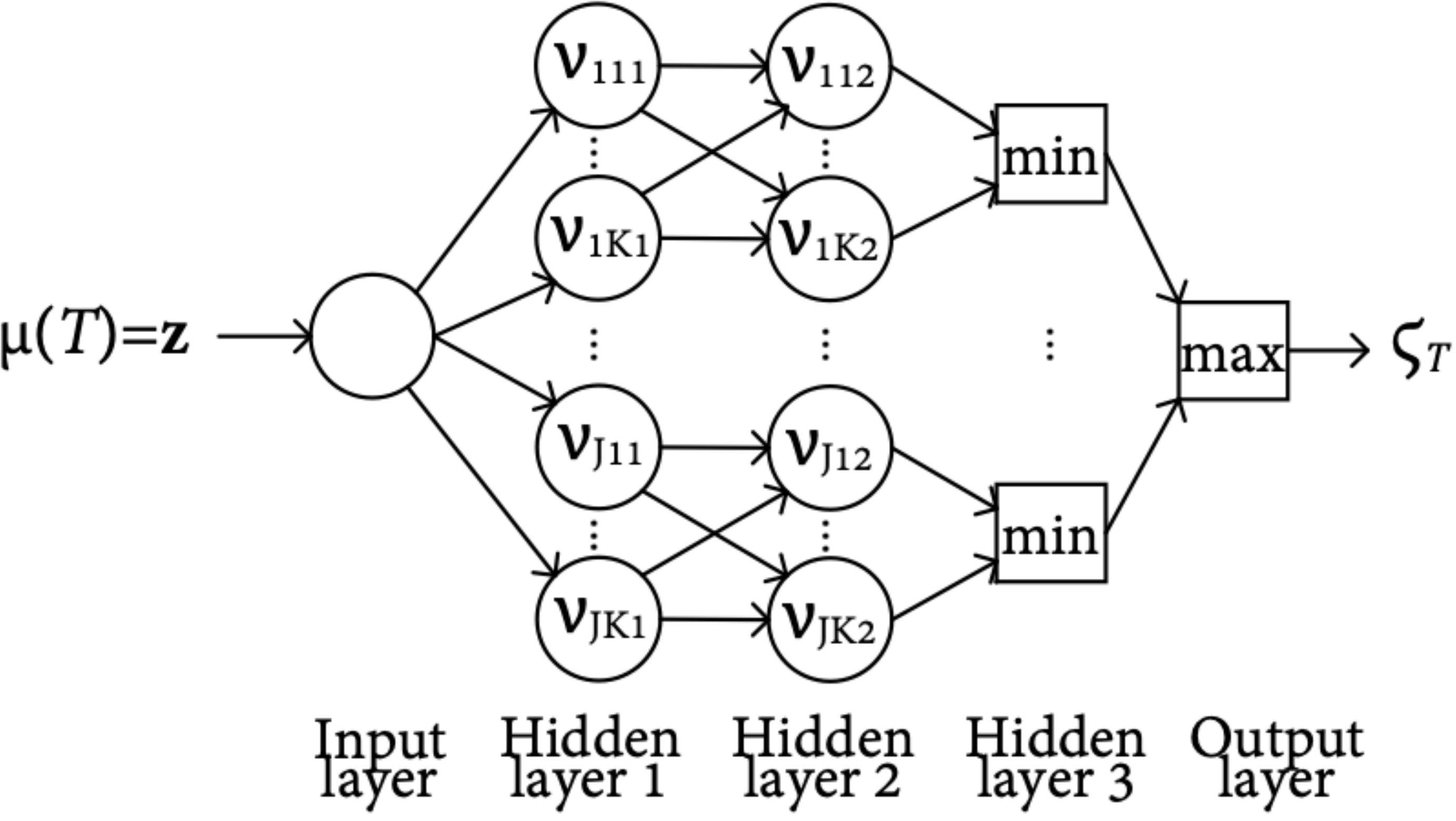}
\par\end{centering}
\caption{Monotonic network $\nu_{\mathbf{w,b}}$ mapping $\mu(\mathcal{T})$
to $\zeta_{\mathcal{T}}$. \label{fig:Monotonic-network}}
\end{figure}
According to Theorem~\ref{thm:Multi-dimensional-generalized-median},
we develop a two-dimensional strategyproof mechanisms by directly
applying Theorem~\ref{thm:(Molin's-max-min-rule)} in each dimension.
We adopt the data preprocessing method in~\cite{Golowich2018}. The
collected location data $\mathbf{x}=(x_{1},\ldots,x_{\hat{N}})$ and
$\mathbf{y}=(y_{1},\ldots,y_{\hat{N}})$ in ascending order, i.e.,
$x_{\pi_{x}(1)}\leq x_{\pi_{x}(2)}\leq\cdots\leq x_{\pi_{x}(\hat{N})}$
and $y_{\pi_{y}(1)}\leq y_{\pi_{y}(2)}\leq\cdots\leq y_{\pi_{y}(\hat{N})}$
where $\pi_{x}(j)$ and $\pi_{y}(j)$ respectively represent the worker
ID at the $j$th place on X and Y axes. Usually, we normalize all
input data into $[0,1]$ in the experiments. We define two sets $\mathcal{T}_{x}(j)=\{\pi_{x}(1),\pi_{x}(2),\ldots,\pi_{x}(j)\}$
and $\mathcal{T}_{y}(j)=\{\pi_{y}(1),\pi_{y}(2),\ldots,\pi_{y}(j)\}$
where $j\in\hat{\mathcal{N}}$. We also establish a monotonically
increasing mapping $\mu(\mathcal{T})$ to transform the set $\mathcal{T}$
to the $\hat{N}$-length binary vector $\mathbf{z}=(z_{1},\ldots,z_{\hat{N}})$
where $z_{i}=1$ if $i\in\mathcal{T}$ and $z_{i}=-1$ if $i\notin\mathcal{T}$.
Thus, if $\mathbf{z}=\mu(\mathcal{T})=(z_{1},\ldots,z_{\hat{N}})$,
$\mathbf{z}'=\mu(\mathcal{T}')=(z'_{1},\ldots,z'_{\hat{N}})$ and
$\mathcal{T}\subseteq\mathcal{T}'$, we can have $z_{i}\leq z'_{i},\forall i\in\hat{\mathcal{N}}$. 

The first condition in Theorem~\ref{thm:(Molin's-max-min-rule)} actually requires a monotonically increasing mapping from a set $\mathcal{T}$ to a constant value $\zeta_{\mathcal{T}}$. As $\mu(\mathcal{T})$ has already mapped the set $\mathcal{T}$ to a vector $\mathbf{z}$, we construct a five-layer neural network $\nu_{\mathbf{w,b}}$ (shown in Fig.~\ref{fig:Monotonic-network}) to approximate a monotonically increasing function, i.e., $\nu_{\mathbf{w,b}}(\mu(\mathcal{T}))=\nu_{\mathbf{w,b}}(\mathbf{z})=\zeta_{\mathcal{T}}$.
The increasing monotonicity here means $\nu_{\mathbf{w,b}}(\mathbf{z})=\zeta_{\mathcal{T}}\leq\nu_{\mathbf{w,b}}(\mathbf{z}')=\zeta_{\mathcal{T}'}$
if $z_{i}\leq z'_{i},\forall i\in\hat{\mathcal{N}}$. The monotonic
neural network function $\nu_{\mathbf{w,b}}$ is described by
\begin{align}
\nu_{\mathbf{w,b}}(\mu(\mathcal{T})) & =\nu_{\mathbf{w,b}}(\mathbf{z})\nonumber \\
 & =\max_{j\in[J]}\min_{k\in[K]}\{s(b_{jk2}+\mathrm{e}^{\mathbf{w}_{jk2}}s(\mathrm{e}^{\mathbf{w}_{j1}}\mathbf{z}^{\mathrm{T}}+\mathbf{b}_{j1}))\},\label{eq:neural-network-func}
\end{align}
where $J$ and $K$ are positive integral hyperparameters that affect the accuracy and complexity of the neural network, $\mathbf{z}^{\mathrm{T}}$ is the transpose of $\mathbf{z}$, $\mathbf{w}_{j1}\in\mathbb{R}^{K\times\hat{N}},\,\mathbf{b}_{j1}\in\mathbb{R}^{K\times1}$ are parameters in the first hidden layer, and $\mathbf{w}_{jk2}\in\mathbb{R}^{1\times K},\,b_{jk2}\in\mathbb{R}$ are the parameters in the second hidden layer. The exponential operations in (\ref{eq:neural-network-func}) are used to guarantee that the weights of the input vector $\mathbf{z}$, i.e., $\mathrm{e}^{\mathbf{w}_{j1}}$ and $\mathrm{e}^{\mathbf{w}_{jk2}}$, are always positive. We use a shifted log-sigmoid function $s(t)=\log(\frac{1}{1+\mathrm{e}^{-t}})+1$ as the activation function which also well restricts the output range. The max-min neural network in Fig.~\ref{fig:Monotonic-network} is monotonically increasing as it follows the characterizations of the monotonic network in~\cite{Sill1998,You2017}. 
\begin{figure}[tbh]
\begin{centering}
\includegraphics[width=0.8\columnwidth]{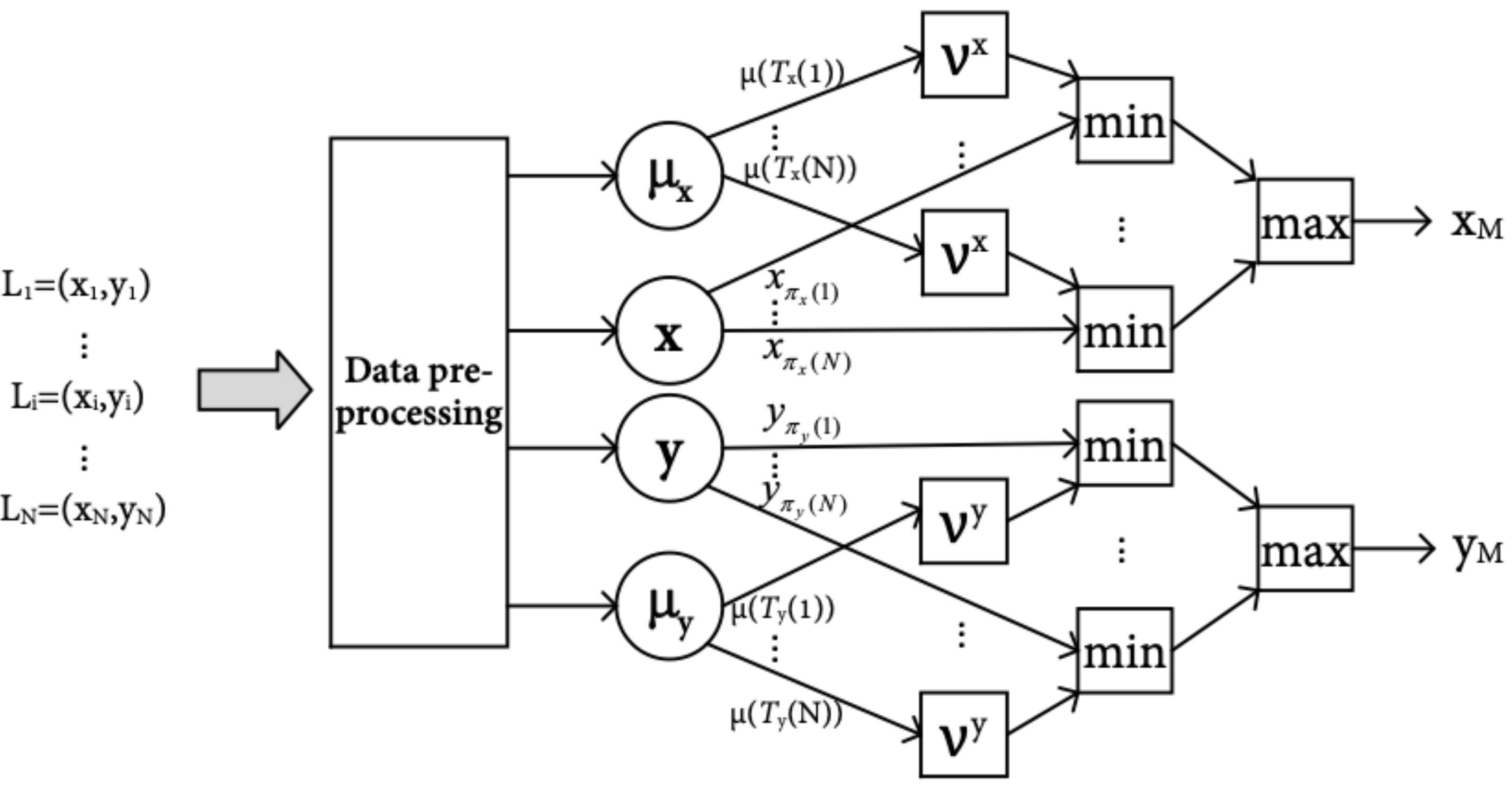}
\par\end{centering}
\caption{The deep neural network $f_{\mathbf{w,b}}$ which forms the MDL mechanism.}
\end{figure}
Next, based on the second condition in Theorem~\ref{thm:(Molin's-max-min-rule)}, we construct the complete deep neural network $f_{\mathbf{w,b}}$ by integrating the monotonic network $\nu_{\mathbf{w,b}}$ with the max and min functions. Finally, the neural network function $f_{\mathbf{w,b}}$ of the MDL mechanism is
\begin{align}
f_{\mathbf{w,b}}(\mu_{x},\mu_{y},\mathbf{x},\mathbf{y}) & =(x_{\mathrm{M}},y_{\mathrm{M}})\nonumber \\
 & =(\max_{i\in\hat{\mathcal{N}}}\left\{ \min\{\nu_{\mathbf{w,b}}^{x}(\mu(\mathcal{T}_{x}(i))),x_{\pi_{x}(i)}\}\right\} ,\nonumber \\
 & \max_{j\in\hat{\mathcal{N}}}\left\{ \min\{\nu_{\mathbf{w,b}}^{y}(\mu(\mathcal{T}_{y}(j)),y_{\pi_{y}(i)}\}\right\} ).\label{eq:f-neural-network-func}
\end{align}
According to Theorems~\ref{thm:Multi-dimensional-generalized-median} and~\ref{thm:(Molin's-max-min-rule)}, the MDL mechanism is strategyproof. Note that the objective function in (\ref{eq:SC_cost_function}) is convex with respect to $L_{\mathrm{M}}=(x_{\mathrm{M}},y_{\mathrm{M}})$. Hence, for each data sample $(\mathbf{x},\mathbf{y})$, we can efficiently compute the optimal solution $L_{\mathrm{M}}^{*}=(x_{\mathrm{M}}^{*},y_{\mathrm{M}}^{*})$ to minimize the SC platform's crowdsourcing cost in (\ref{eq:SC_cost_function}) without considering strategyproofness and then use it as the label. In the training process, we adopt the mean squared error (MSE) to evaluate the training loss and optimize the deep neural network parameters. Given a set of $G$ data samples $\mathcal{G}=\{(\mathbf{x},\mathbf{y})^{1},\ldots,(\mathbf{x},\mathbf{y})^{G}\}$ and corresponding labels $\mathbf{L}_{\mathrm{M}}^{*}=\{(x_{\mathrm{M}}^{*},y_{\mathrm{M}}^{*})^{1},\ldots,(x_{\mathrm{M}}^{*},y_{\mathrm{M}}^{*})^{G}\}$, the loss can be calculated by 
\begin{align}
loss & =\frac{1}{G}\sum_{j=1}^{G}(\hat{l}_{m}(\mathsf{M}_{\mathrm{MDL}}((\mathbf{x},\mathbf{y})^{j});(\mathbf{x},\mathbf{y})^{j})\nonumber \\
 & -\hat{l}_{m}((x_{\mathrm{M}}^{*},y_{\mathrm{M}}^{*})^{j};(\mathbf{x},\mathbf{y})^{j}))^{2},\label{eq:Loss}
\end{align}
where $\mathsf{M}_{\mathrm{MDL}}((\mathbf{x},\mathbf{y})^{j})$ is the mobile BS's location output by the MDL mechanism when the input is the $j$th data sample $(\mathbf{x},\mathbf{y})^{j},j\in\{1,\ldots,G\}$. 

\section{Experimental results and discussions\label{sec:Experimental-and-simulation} }

In this section, we conduct simulations based on real data to evaluate the performance of our proposed framework and strategyproof deployment mechanisms. Unless otherwise stated, the simulation configuration is set as follows. We consider a $[0,200]\times[0,200]$ square-meter area as the SC task area $\mathrm{\boldsymbol{A}}_{t}$. The number of registered workers is set at $N=40$. We set the height of the mobile BS $h=10\,\mathrm{m}$, e.g., a drone, the channel gain to noise ratio $g=90\,\mathrm{dB}$, the bandwidth of each subchannel $B=60\,\mathrm{MHz}$, the data utility parameters $a_{1}=10^{4}$, $a_{2}=200$, the energy conversion efficiency $\eta=0.6$, the antenna gain $\Gamma=-30\,\mathrm{dB}$, and the path-loss exponent $\alpha=2$~\cite{Ju2014}. The sensing energy cost per bit $b_{i}$ is generated from the uniform distribution on $[10^{-4},1.1\times10^{-4}]$. Each measurement is averaged over more than $100$ instances. 

\begin{figure}[tbh]
\begin{centering}
\includegraphics[width=0.8\columnwidth]{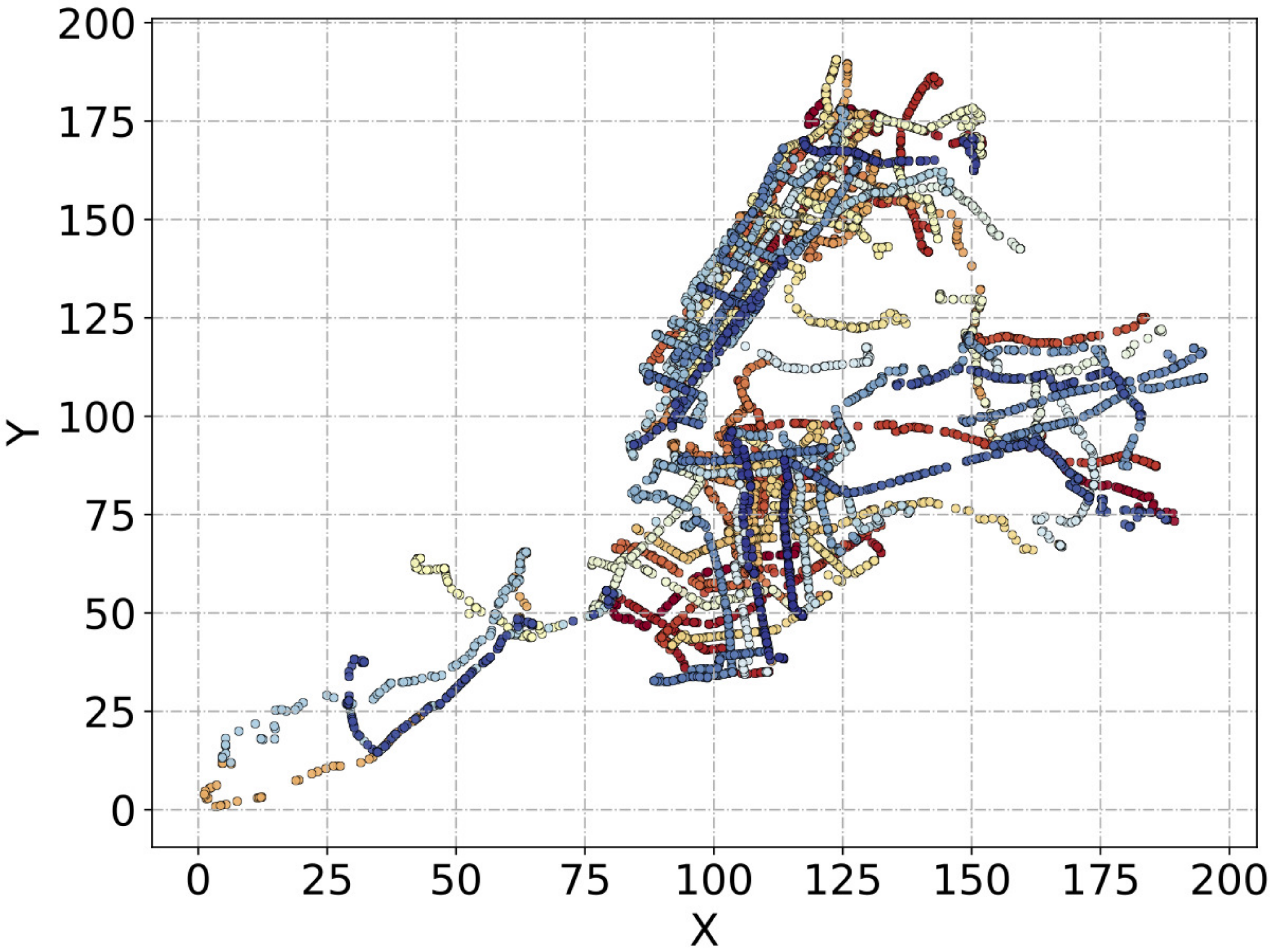}
\par\end{centering}
\caption{A brief overview of the prepared bus mobility dataset (each color
represents a worker).\label{fig:dataset_overview}}
\end{figure}

To illustrate the practical use of our proposed algorithms, we use a real-world dataset from NYC MTA Real-Time Data Feeds\footnote{ https://datamine.mta.info/}. The dataset has more than $2$ million mobility traces, i.e. the GPS location records, of $95$ workers located in New York City over a period of one month. It is reasonable that a worker usually estimates the working area according to its past experience. Therefore, the historical GPS records help us to calculate the worker's working area $\boldsymbol{A}_{i}$ and maximum distance $D_{i}$. For better performance of neural network processing, we first normalize the dataset to the range $[0,1]$ and respectively prepare $24,000$ samples (training dataset) for MDL model training and $6,000$ samples (testing dataset) for testing and performance evaluation. Each data sample contains the workers' locations at one time slot. We randomly choose $100$ samples to provide a brief overview of the prepared dataset, as shown in Fig.~\ref{fig:dataset_overview}. Each worker's maximum distance $D_{i}$ is also calculated according to the dataset. We use the Pytorch deep learning library to implement the MDL mechanism with $K=8,\,J=8$. We use the ADAM optimizer with a learning rate of $0.005$ and mini-batch of $200$ when training the MDL model. All the experiments were run on a workstation with a GTX1080Ti GPU. 

\begin{figure}[tbh]
\begin{centering}
\includegraphics[width=0.8\columnwidth]{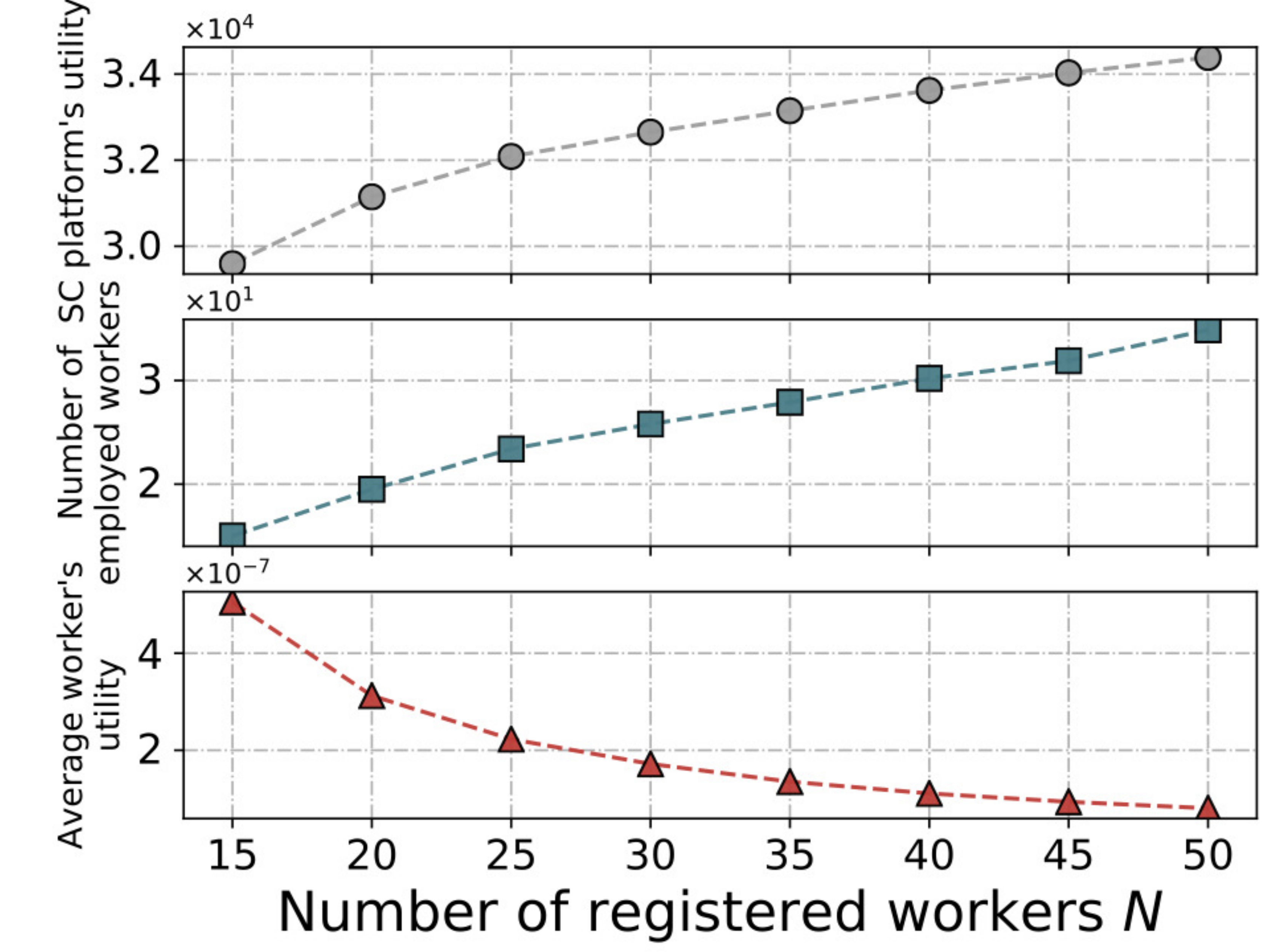}
\par\end{centering}
\begin{centering}
\caption{Impact of the number of registered workers.\label{fig:Impact_of_Ne}}
\par\end{centering}
\end{figure}
\begin{figure}[tbh]
\begin{centering}
\includegraphics[width=0.8\columnwidth]{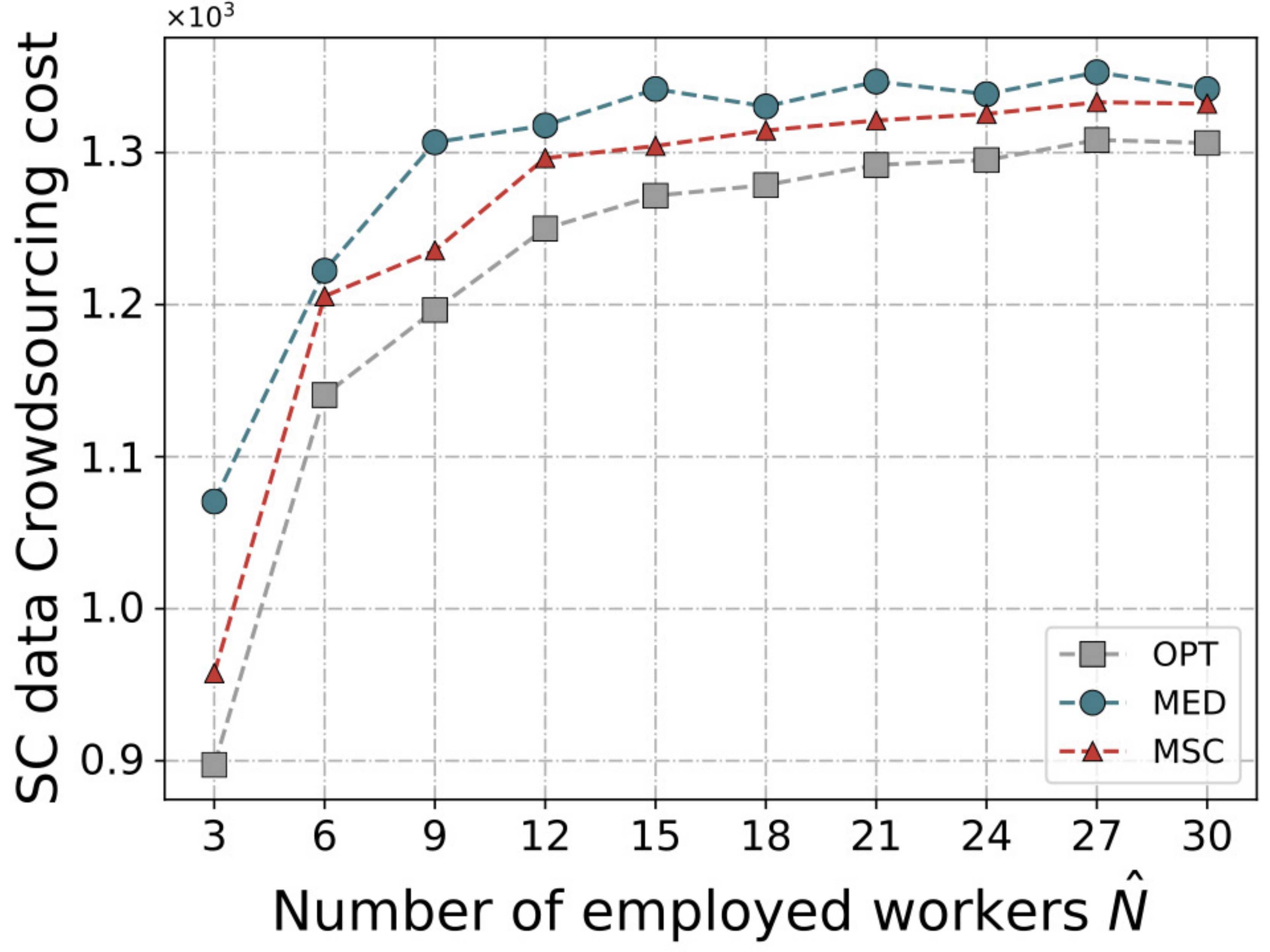}
\par\end{centering}
\caption{The SC data crowdsourcing cost achieved by different mechanisms with
varied number of employed workers $\hat{N}$ in the special case ($\alpha=2$).\label{fig:SC_utility_specialcase}}
\end{figure}
Figure~\ref{fig:Impact_of_Ne} demonstrates the impact of the number of registered workers $N$ on the SC platform's utility, the average worker's utility and the number of employed workers in the task allocation phase. When the number of registered workers increases, the SC platform's utility and the number of employed workers gradually increase but with a diminishing return. These reflect that when more workers are employed, the SC platform has to consume more charging power for the same marginal utility. By contrast, the average worker's utility decreases with the increase of registered workers because of the more competition among workers. Next, we present simulation results for the data crowdsourcing phase. 

Figure~\ref{fig:SC_utility_specialcase} depicts the performance of the proposed truthful MSC mechanism in the special case as discussed in Section~\ref{subsec:Conventional-mechanism}. As a priori information, the workers' locations are i.i.d. uniformly distributed over the SC task area. Thus, the added single constant point $(x_{\mathrm{c}},y_{\mathrm{c}})$ is set at the expected location $(100,100)$ due to the symmetry and the analysis presented in Section~\ref{subsec:Conventional-mechanism}. The optimal solution without considering the incentive constraints is also calculated for comparison, which is denoted as the OPT algorithm. Obviously, the performance of the MSC mechanism is better (with lower crowdsourcing cost) than that of the MED mechanism when $\hat{N}=3$, which is consistent with the theoretical analysis. For $\hat{N}>3$, the MSC mechanism still outperforms the MED mechanism but is always inferior to the OPT mechanism because of the sacrifice for guaranteeing the strategyproofness. 

To illustrate the performance of our proposed mechanisms in minimizing the SC data crowdsourcing cost $\hat{l}_{m}$, we use the \emph{average performance ratio} $\omega^{\mathrm{avg}}$ and the \emph{worst-case} \emph{performance ratio} $\omega^{\mathrm{wst}}$ as the evaluation metrics. In our experiment, they are measured based on the prepared test dataset. The average performance ratio is defined as the ratio of the average data crowdsourcing cost achieved by the proposed mechanism over the average crowdsourcing cost achieved by the OPT mechanism. The worst-case performance ratio is defined as the highest ratio of the data crowdsourcing cost achieved by the proposed mechanism over the crowdsourcing cost achieved by the OPT mechanism. Formally, given the test dataset of $G_{\mathrm{test}}$ data samples $\mathcal{G}_{\mathrm{test}}=\{(\mathbf{x},\mathbf{y})^{1},\ldots,(\mathbf{x},\mathbf{y})^{G_{\mathrm{test}}}\}$, we take the MED mechanism for example and have $\omega_{\mathrm{MED}}^{\mathrm{avg}}=\frac{\frac{1}{G_{\mathrm{test}}}\sum_{(\mathbf{x},\mathbf{y})^{j}\in\mathcal{G}_{\mathrm{test}}}\hat{l}_{m}(\mathsf{M}_{\mathrm{MED}}((\mathbf{x},\mathbf{y})^{j});(\mathbf{x},\mathbf{y})^{j})}{\frac{1}{G_{\mathrm{test}}}\sum_{(\mathbf{x},\mathbf{y})^{j}\in\mathcal{G}_{\mathrm{test}}}\hat{l}_{m}(\mathsf{M}_{\mathrm{OPT}}((\mathbf{x},\mathbf{y})^{j});(\mathbf{x},\mathbf{y})^{j})}$
and $\omega_{\mathrm{MED}}^{\mathrm{wst}}=\max_{(\mathbf{x},\mathbf{y})^{j}\in\mathcal{G}_{\mathrm{test}}}\frac{\hat{l}_{m}(\mathsf{M}_{\mathrm{MED}}((\mathbf{x},\mathbf{y})^{j});(\mathbf{x},\mathbf{y})^{j})}{\hat{l}_{m}(\mathsf{M}_{\mathrm{OPT}}((\mathbf{x},\mathbf{y})^{j});(\mathbf{x},\mathbf{y})^{j})}$. A lower ratio means a better performance. 

\begin{figure}[tbh]
\begin{centering}
\includegraphics[width=0.8\columnwidth]{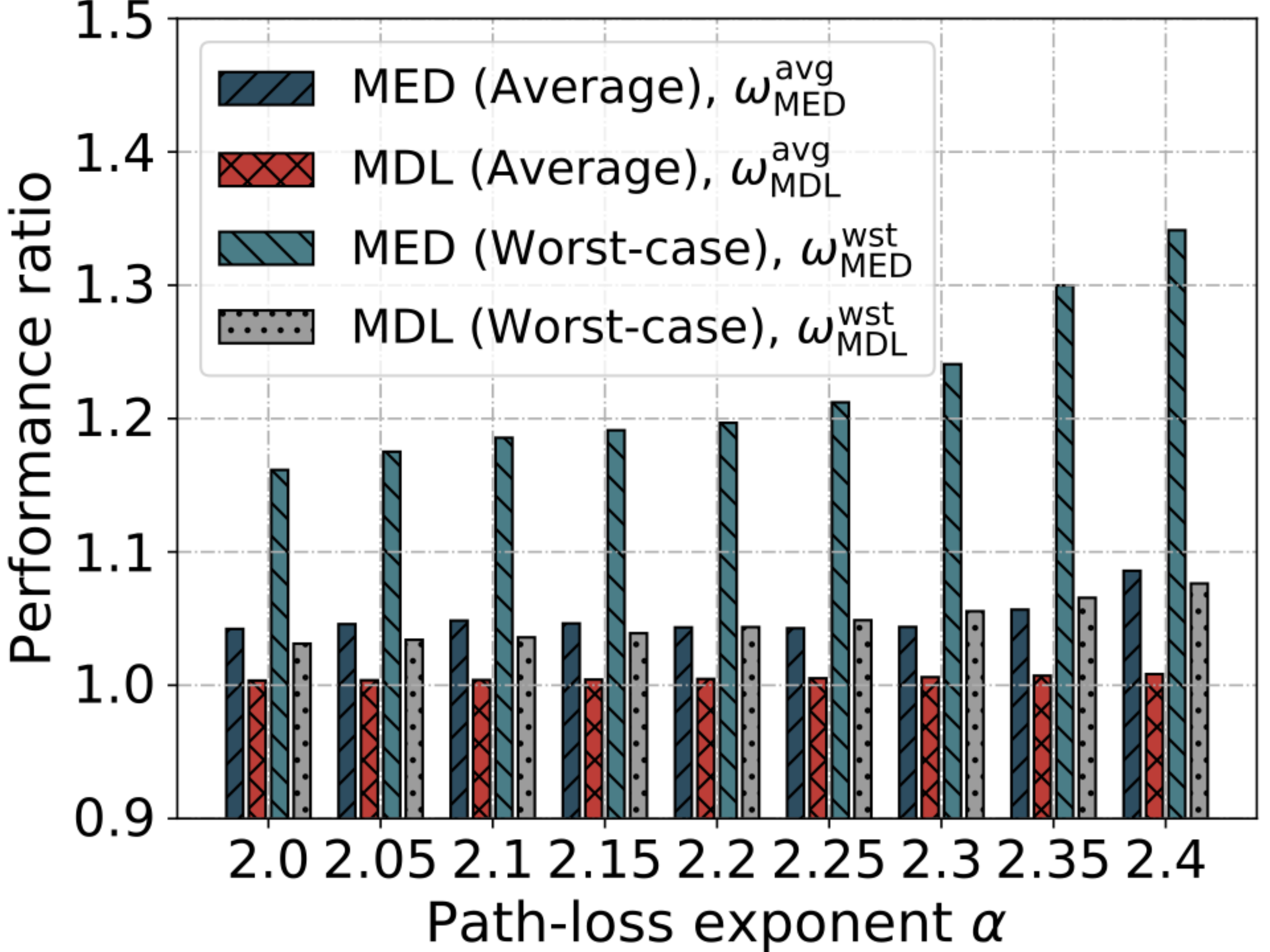}
\par\end{centering}
\caption{The performance ratio with varied path-loss exponent.\label{fig:ratio_alpha}}
\end{figure}
\begin{figure}[tbh]
\begin{centering}
\includegraphics[width=0.8\columnwidth]{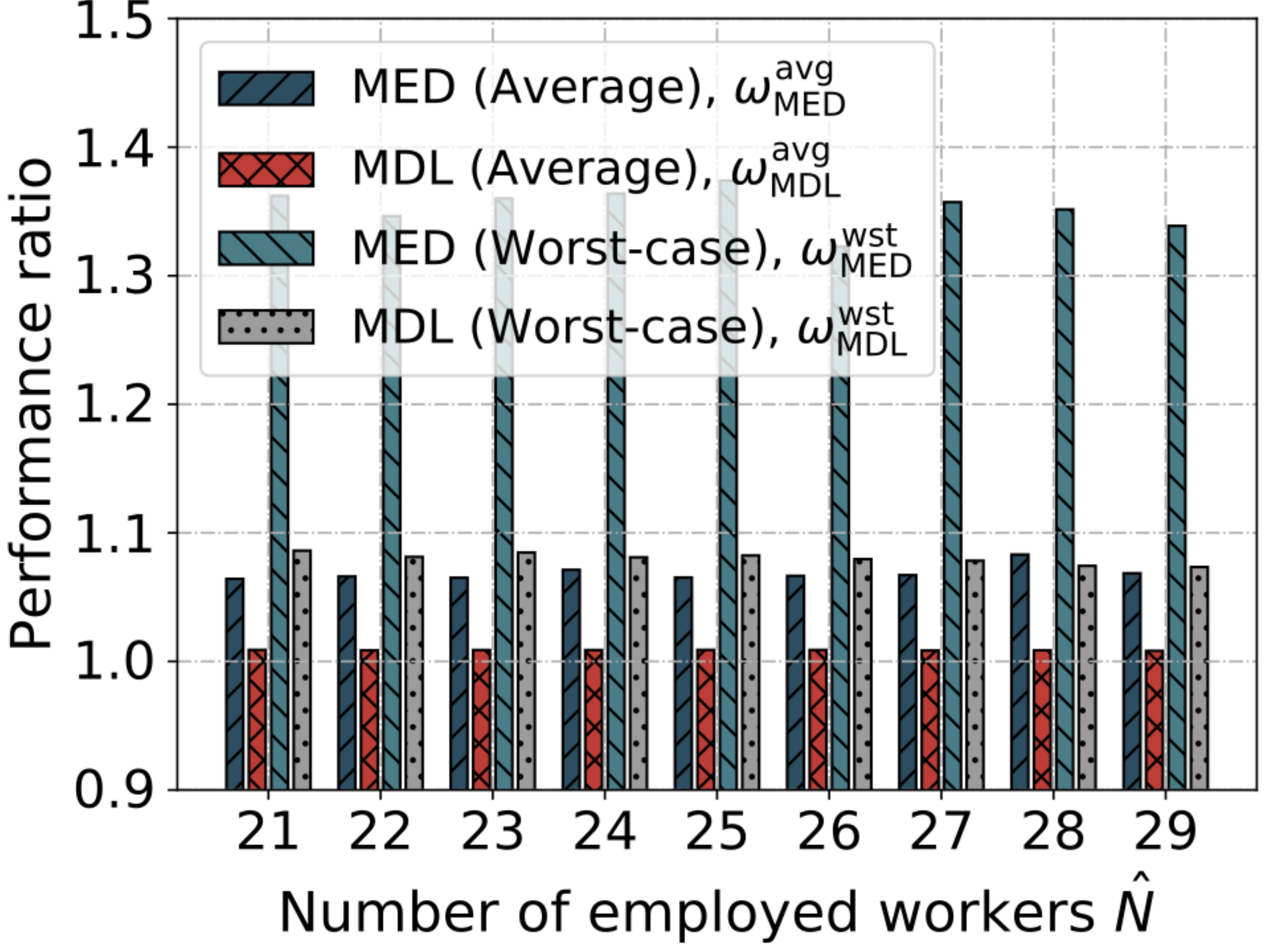}
\par\end{centering}
\caption{The performance ratio with varied number of employed workers.\label{fig:ratio_NEw}}
\end{figure}
\vspace{-0.2cm}

In Fig.~\ref{fig:ratio_alpha}, the number of employed workers $\hat{N}$ is fixed to be $30$ and we investigate the performance of the MED mechanism and the MDL mechanism with the varied path\textendash loss exponent. We find that when the radio environment gets worse (a larger path\textendash loss exponent $\alpha$), the average and worst-case performance ratios of both the MED and the MDL mechanism grow at different rates. In Fig.~\ref{fig:ratio_NEw}, we fix the path-loss exponent $\alpha$ at $2.4$ and study the impact of the different number of employed workers on the performance ratios of each proposed mechanism. Figure~\ref{fig:ratio_NEw} illustrates that the increasing number of employed workers has an implicit impact on the performances of both proposed mechanisms. The main reason is that each worker's location distribution in the mobility dataset is different. Otherwise, if each worker's location follows the i.i.d distribution, more employed workers mean more reported data which makes the hidden distribution more certain and at least makes the average performance ratio of the MED mechanism decline. This phenomenon can be seen in Fig.~\ref{fig:SC_utility_specialcase}. Therefore, the impact of the number of employed workers is closely related to the characteristic of the used dataset. In summary, compared with the MED mechanism, the deep learning based mechanism, i.e., the MDL mechanism, shows two clear advantages in the considered complicated scenario. The first advantage is noticeable stability. In Fig.~\ref{fig:ratio_alpha}, it can be observed that the worst performance ratio of the MED mechanism increases exponentially with the increasing path-loss exponent, while the MDL mechanism shows an approximately linear increasing trend. The second advantage is the significant performance improvement. As illustrated in Fig.~\ref{fig:ratio_NEw}, the MDL mechanism achieves at least $5.19$\% ($18.39$\%) reduction in average (worst-case) performance ratio compared to the MED mechanism.
\vspace{-0.5cm}

\section{Conclusion\label{sec:Conclusion}}

In this paper, we have proposed a wireless powered spatial crowdsourcing framework composed of two phases. In the task allocation phase, we have proven that the proposed Stackelberg game based incentive mechanism can help the SC platform efficiently allocate the tasks and the wireless charging power. For the deployment of the mobile BS in the data crowdsourcing phase, we have adopted the classical strategyproof median mechanism. We have also designed a conventional strategyproof mechanism and a deep learning based strategyproof mechanism from a Bayesian point of view. Besides avoiding the dishonest worker's manipulation, extensive experimental results based on synthetic and real-world datasets demonstrate the effectiveness of the proposed framework in allocating tasks and charging power to workers. It is worth noting that, in this paper, we use the data transmission rate as a general metric to evaluate the data utility. In future work, we will take other specific data attributes into consideration, such as the location and timeliness. For the time-varying scenario, we can also develop a dynamical deployment mechanism using deep reinforcement learning. 

\bibliographystyle{IEEEtran}
\bibliography{IEEEabrv,PJ6_TVT_WPCrowdsourcing_deeplearning}

\end{document}